\documentclass[prd,preprint,tightenlines,floatfix,showpacs,preprintnumbers,nofootinbib,eqsecnum]{revtex4}

 \usepackage[dvips,final]{graphicx}
  \usepackage{amssymb}
   \usepackage{amsmath}
    \usepackage{amsfonts}
     \usepackage{epsfig}
      \usepackage{bm}

\begin{document}

\thispagestyle{empty} \preprint{\hbox{}} \vspace*{-10mm}

\title{Central exclusive production of scalar $\chi_c$ meson
\\at the Tevatron, RHIC and LHC energies}

\author{R.~S.~Pasechnik}
\email{rpasech@theor.jinr.ru}

\author{A.~Szczurek}
\email{antoni.szczurek@ifj.edu.pl}

\author{O.~V.~Teryaev}
\email{teryaev@theor.jinr.ru}

\affiliation{ Bogoliubov Laboratory of Theoretical Physics, JINR,
Dubna 141980, Russia}

\affiliation{Institute of Nuclear Physics PAN, PL-31-342 Cracow,
Poland} \affiliation{University of Rzesz\'ow, PL-35-959 Rzesz\'ow,
Poland}

\date{\today}

\begin{abstract}
We calculate several differential distributions for exclusive
double diffractive $\chi_c(0^{++})$
production in proton-antiproton collisions at the Tevatron and in
proton-proton collisions at RHIC and LHC in terms of unintegrated gluon
distributions (UGDFs) within the $k_t$-factorisation approach.
The uncertainties of the Khoze-Martin-Ryskin approach are discussed
in detail. The $g^* g^* \to \chi_c(0^{++})$
transition vertex is calculated as a function of gluon
virtualities applying the standard pNRQCD technique.
The off-shell effects are discussed and quantified. They
lead to a reduction of the cross section by a factor 2--5, depending
on the position in the phase space and UGDFs.
Different models of UGDFs are used and the results are shown and discussed.
The cross section for diffractive component depends strongly on UGDFs.
We calculate also the differential distributions for
the $\gamma^* \gamma^* \to \chi_c(0^{++})$ fusion mechanism.
The integrated cross section for photon-photon fusion is
much smaller than that of diffractive origin. The two components have
very different dependence on momentum transfers $t_1, t_2$ in the
nucleon lines as well as azimuthal-angle correlations between both
outgoing nucleons.
\end{abstract}


\maketitle

\section{Introduction}

The discovery of Higgs is the main motivation for the construction
and putting into operation the Large Hadron Collider (LHC) at
CERN. The analysis of inclusive cross section will be the ``main
road'' of the future investigations. Different decay channels will
be studied. The analysis in each (!) of these channels is rather
complicated as huge unreduceable backgrounds are unavoidably
present.

\begin{figure}[!h]    
 \centerline{\includegraphics[width=0.4\textwidth]{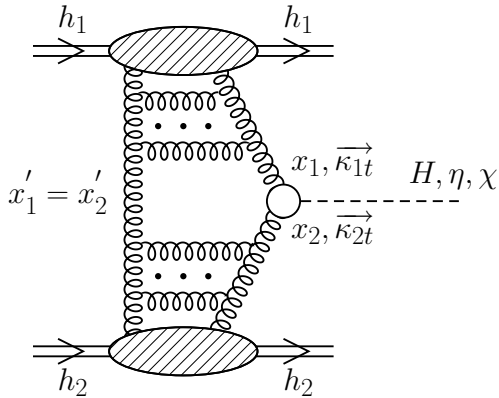}}
   \caption{\label{fig:QCDdiff}
   \small  The sketch of the bare QCD mechanism. The kinematical
variables are shown in addition.}
\end{figure}

The diffractive exclusive production of Higgs boson seems to be
much cleaner in this respect. Many estimates of the corresponding
cross section has been presented in the literature. The so-called
Durham model \cite{KMR} is the state-of-art in this field. The
cross section for the diffractive production is much smaller than
the cross section for the inclusive case, but the ratio of the
signal to more conventional background seems promising.
Recently appeared a detailed analysis of diffractive production
of MSSM Higgs \cite{KMR_MSSM}.
We do not need to mention that any check of the used theoretical methods
against experimental data are not possible at present, at least
for the Higgs production. The way out is to study the diffractive
production of heavy quarkonia where almost the same theoretical
methods can be used. The basic diagram for the amplitude of the
process is shown in Fig.~\ref{fig:QCDdiff}. The production of
heavy quarkonia received a lot of attention from both theory and
experiment in recent years. For a review we refer to
Refs.\cite{Sch94,BF,Bottom,Kramer01,Brambilla}.

QCD dynamics at small quark and gluon momentum fractions (or large
total energy), relevant for HERA, Tevatron, RHIC
and LHC, is still poorly understood. It was shown in
\cite{HKSST00,LSZ02} that the combination of the
$k_{\perp}$-factorisation approach \cite{CCH90,CE91,CC96,RSS99}
and the next-to-leading-logarithmic-approximation (NLLA) BFKL
vertex \cite{FL96} gives quite good agreement with data on
inclusive $Q\bar{Q}$-production. One can therefore hope that these
concepts provide a valueable foundation also for exclusive
processes.

Kaidalov, Khoze, Martin and Ryskin proposed to calculate
diffractive double elastic \footnote{Both protons survive the
collision.} production of Higgs boson in terms of unintegrated
gluon distributions \cite{KMR}. It is not clear at present how
reliable such calculations are, and it would be interesting to
apply this approach to heavy quarkonia production.
In Refs.~\cite{Yuan01,KMR-chi} integrated
cross section for exclusive double diffractive $\chi_c(0^{++})$
production was estimated with the identical formulae as for
the scalar Higgs production with $\Gamma_{H \to gg}$ replaced by
$\Gamma_{\chi_c(0^{++}) \to gg}$. Of course, such a procedure can
be strictly right in general case only for ficticious
structureless objects, when the internal wave function and gluon
virtualities are neglected. From the spectroscopy point of view
the $\chi_c(0^{++})$ meson is a quark-antiquark P-wave state, and
it might be interesting to study an exclusive production of
P-waves applying pNRQCD methods. Such a calculation may be
especially important when we go to larger gluon virtualities.

Parallel to the exclusive channel studies there was a lot of
theoretical activity for inclusive charmonium and bottomonium
production; see for example \cite{HKSST01,Vasin,Likhoded}. There
non-relativistic pQCD methods are usually applied. In these
approaches the nonrelativistic quark-antiquark wave function is
taken explicitly into the calculation. The vertex function
$g^*g^*q\bar q$ corresponds to so-called quasi multi-Regge
kinematics (QMRK), i.e. when $q$ and $\bar{q}$ have similar
rapidities and form a cluster. It is based on the formalism
developed by Lipatov and Fadin \cite{FL96} for diffractive $q\bar
q$ pair production.

\begin{figure}[!h]    
 \centerline{\includegraphics[width=0.3\textwidth]{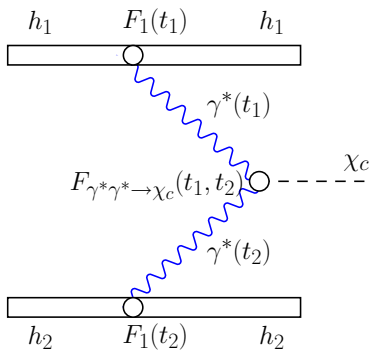}}
   \caption{\label{fig:gamgam}
   \small  The sketch of the photon-photon fusion mechanism.
Form factors appearing in different vertices are shown
explicitly.}
\end{figure}

In the present work we shall use the pNRQCD methods for exclusive
double diffractive $\chi_c(0^{++})$ production. The main
ingredients of our calculations are the unintegrated gluon
distributions (UGDFs) and the effective NLLA BFKL production
vertex in QMRK (with reggeised gluon couplings to quarks). We
would like to compare results obtained using this vertex with the
KMR results. The projection of the heavy quark-antiquark pair onto
the corresponding charmonium state is described in the standard
way within the non-relativistic-quarkonium-model
\cite{Guberina,Baier,Cho-Leibo}.
Finally, we shall refer to the KMR result \cite{KMR-chi} where the wave
function is not included explicitly. For completeness, we shall
include also photon-photon fusion mechanism of exclusive
$\chi_c(0^{++})$ production shown in Fig.~\ref{fig:gamgam}. In
addition to the diffractive QCD approach we discuss also a phenomenological
Pomeron-Pomeron fusion approach.

In the present paper we wish to calculate differential distributions
for exclusive $\chi_c(0^+)$ production with different UGDFs from
the literature. We shall calculate matrix elements for off-shell gluons.
We shall discuss also uncertainty related to the approximation made,
to the choice of the scale, etc. The contribution of the $\gamma^* \gamma^*$
fusion to the differential cross sections will be calculated.

\section{Diffractive QCD mechanism}

\begin{figure}[h]    
\centerline{\epsfig{file=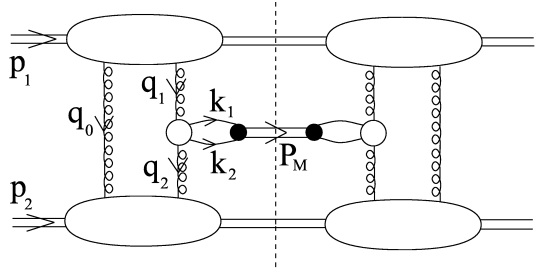,width=7cm}} \caption{The
basic diagram of double-diffractive charmonium production
$pp\rightarrow pp\chi_c$} \label{fig:cut}
\end{figure}

In the pNRQCD approach the diffractive exclusive reaction is
viewed as shown in Fig.~\ref{fig:cut}. The corresponding
calculation can be summarized as follows. First, the
$q\bar{q}$-continuum amplitude is calculated. Then the $gg\to
q\bar{q}$ amplitude is reduced to the $gg\to \chi_c$ amplitude with
standard projection techniques developed in \cite{Guberina,Kuhn}.

\subsection{General kinematics of double diffractive
QCD mechanism}

\begin{figure}[h!]    
\centerline{\epsfig{file=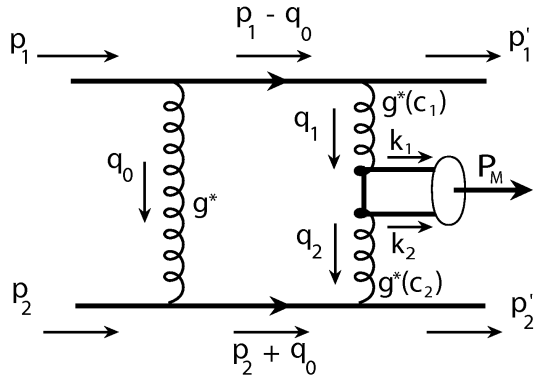,width=7cm}}
\caption{General kinematics of exclusive double-diffractive
production.} \label{fig:kinematics_qcd}
\end{figure}

Let's consider first the $q\bar{q}$-production. The kinematic
variables for the process $pp\rightarrow pp(q\bar{q})$ on the
quark level is shown in Fig.~\ref{fig:kinematics_qcd}.

We use the following definition of the light cone coordinates
\[
k^{+}\equiv n^+_{\alpha}k^{\alpha}=k^{0}+k^{3},\text{ }k^{-}\equiv
n^-_{\alpha}k^{\alpha}=k^{0}-k^{3},\text{ }k_{t
}=(0,k^{1},k^{2},0)=(0,{\bf k,}0{\bf )},
\]
where $n^{\pm}$ are the light-cone basis vectors. In the c.m.s.
frame
\begin{eqnarray}
\label{lcbasis} n^+=\frac{p_2+q_0}{E_{cms}},\qquad
n^-=\frac{p_1-q_0}{E_{cms}},
\end{eqnarray}
and the momenta of the scattering hadrons are given by
\[
p_{1}^{+}=p_{2}^{-}=\sqrt{s},\text{ \ }p_{1}^{-}=p_{2}^{+}=p_{1,t
}=p_{2,t }=0,
\]
where the Mandelstam variable $s=4E_{cms}^2.$ The momenta of the
t-channel gluons are $q_0$, $q_{1}$ and $q_{2}$ (see
Fig.~\ref{fig:kinematics_qcd}). The on-shell quark and antiquark
(with mass $m$) have momentum $k_{1}$ respectively $k_{2}$ with
\[
k_{1}^{-}=\frac{(m^{2}-k_{1,t }^{2})}{k_{1}^{+}},\text{ \ }k_{2}^{-}=%
\frac{(m^{2}-k_{2,t }^{2})}{k_{2}^{+}}.
\]
In the high energy (large $s$) regime we have
\begin{eqnarray*}
&&P^{+}\!=q_{1}^{+}-q_{2}^{+}\approx q_{1}^{+}, \;
P^{-}\!=q_{1}^{-}-q_{2}^{-}\approx -q_{2}^{-},
\end{eqnarray*}
where $P=k_{1}+k_{2}$ is the momentum of the heavy charmonium with
mass $M$: $P^{2}=M^2\simeq4m^2$. The longitudinal momentum
fractions of the gluons are $x_{1}=q_{1}^{+}/p_{1}^{+}$,
$x_{2}=-q_{2}^{-}/p_{2}^{-}$. Of course, we will finally assume
that $q_{0/1/2,t}^2=-|{\bf q}_{0/1/2,t}|^2.$

The decomposition of gluon momenta into longitudinal and
transverse parts gives
\begin{eqnarray}
&&q_1=x_1(p_1-q_0)+q_{1,t},\qquad q_2=-x_2(p_2+q_0)+q_{2,t} \\
&& 0<x_{1,2}<1,\qquad
q_0=x'_1p_1+q_{0,t}=-x'_2p_2+q_{0,t}\,.\nonumber\label{dec}
\end{eqnarray}
We take into account below that $x'_1=x'_2=x_0$. Making use of
conservation laws we get
\begin{eqnarray}
q_1+p'_1=p_1-q_0,\qquad q_2+p_2+q_0=p'_2\, . \label{CL}
\end{eqnarray}
Taking the transverse parts from these relations gives
\begin{eqnarray}
q_{1,t}=-(p'_{1,t}+q_{0,t}),\qquad q_{2,t}=p'_{2,t}-q_{0,t}\, .
\label{perp}
\end{eqnarray}
%

\subsection{Matrix element for exclusive double diffractive
$\chi_c(0^{++})$ production}

According to Khoze-Martin-Ryskin approach (KMR) \cite{KMR}, we
write the amplitude of exclusive double diffractive colour singlet
production $pp\to pp\chi_{cJ}$ as
\begin{eqnarray}
{\cal
M}^{g^*g^*}=\frac{s}{2}\cdot\pi^2\frac12\frac{\delta_{c_1c_2}}{N_c^2-1}\,
\Im\int
d^2 q_{0,t}V^{c_1c_2}_J\frac{f^{off}_{g,1}(x_1,x_1',q_{0,t}^2,
q_{1,t}^2,t_1)f^{off}_{g,2}(x_2,x_2',q_{0,t}^2,q_{2,t}^2,t_2)}
{q_{0,t}^2\,q_{1,t}^2\, q_{2,t}^2}. \label{ampl}
\end{eqnarray}
The normalization of this amplitude differs from the KMR one
\cite{KMR-chi,KMR} by the factor $s/2$ and coincides with the
normalization in our previous work on exclusive $\eta'$-production
\cite{SPT07}. The amplitude is averaged over the colour indices
and over two transverse polarisations of the incoming gluons
\cite{KMR}. The bare amplitude above is subjected to absorption
corrections which depend on collision energy. We shall discuss
this issue shortly when presenting our results.

The vertex factor $V_J^{c_1c_2}=V_J^{c_1c_2}(q_{1,t}^2,
q_{2,t}^2,P_{Mt}^2)$ in expression (\ref{ampl}) describes the
coupling of two virtual gluons to $\chi_{cJ}$-meson that can be
written as
\begin{eqnarray}\label{vert}
V^{c_1c_2}_J={\cal
P}(q\bar{q}\rightarrow\chi_{cJ})\bullet\Psi^{c_1c_2}_{ik}(k_1,k_2) \; ,
\end{eqnarray}
where ${\cal P}(q\bar{q}\rightarrow\chi_{cJ})$ is the operator
that projects the $q\bar{q}$ pair onto the charmonium bound state
(see below), $\Psi^{c_1c_2}(k_1,k_2)$ is the production amplitude
of a pair of massive quark $q$ and antiquark $\bar{q}$ with
momenta $k_1$, $k_2$, respectively.

\begin{figure}[h!]
\centerline{\epsfig{file=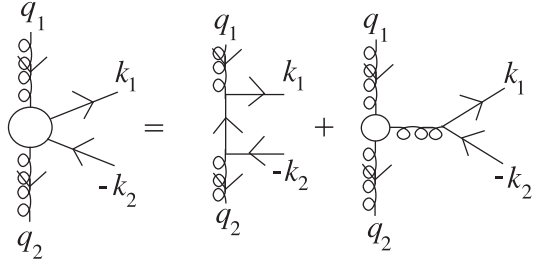,width=7cm}} \caption{The
effective vertex within QMRK} \label{flvertex}
\end{figure}

Within the QMRK approach \cite{FL96} we have
\begin{eqnarray}\label{qqamp}
\Psi(c_1,c_2;i,k;k_1,k_2)&=&
-g^2(t^{c_1}_{ij}t^{c_2}_{jk}b(k_1,k_2)-t^{c_2}_{kj}t^{c_1}_{ji}\bar{b}(k_2,k_1)),\quad
\alpha_s=\frac{g^2}{4\pi} \; ,
\end{eqnarray}
where $t^c$ are the colour group generators in the fundamental
representation, $b,\,\bar{b}$ are the effective vertices
(\ref{bb}) arising from the Feynman rules of QMRK illustrated in
Fig.~(\ref{flvertex}).

%
\begin{eqnarray} \label{bb}
b(k_1,k_2)=\gamma^-\frac{\hat{q}_{1}-\hat{k}_{1}-m}{(q_1-k_1)^2-m^2}
\gamma^+-\frac{\gamma_{\beta}{\Gamma^{+-}}^{\beta}(q_2,q_1)}{(k_1+k_2)^2}
\; , \\
\bar{b}(k_1,k_2)=\gamma^+\frac{\hat{q}_{1}-\hat{k}_{1}+m}{(q_1-k_1)^2-m^2}
\gamma^--\frac{\gamma_{\beta}{\Gamma^{+-}}^{\beta}(q_2,q_1)}{(k_1+k_2)^2}
\; .
\nonumber
\end{eqnarray}
Taking into account definitions (\ref{lcbasis}) and (\ref{dec}) and
using the gauge invariance property
\begin{eqnarray*}
q_1^{\nu}V^{c_1c_2}_{J,\,\mu\nu}=q_2^{\mu}V^{c_1c_2}_{J,\,\mu\nu}=0
\end{eqnarray*}
we get the following projection
\begin{eqnarray} \label{decomp}
&&{}V^{c_1c_2}_{J}=
n^+_{\mu}n^-_{\nu}V_{J,\,\mu\nu}^{c_1c_2}(q_1,q_2)=-
\frac{4}{s}\frac{q^{\nu}_{1,t}}{x_1}\frac{q^{\mu}_{2,t}}{x_2}
V^{c_1c_2}_{J,\,\mu\nu}(q_1,q_2).
\end{eqnarray}
Normalization of the polarization vectors coincides with the one
in Ref.~\cite{Forshaw05}. Since we adopt here the definition of
the polarization vectors proportional to gluon transverse momenta
$q_{1/2,t}$, then we must take into account the longitudinal
momenta in numerators of vertices (\ref{bb}). While projecting on
the color singlet the $ggg$-verticies ${\Gamma^{+-}}^{\beta}$ in
(\ref{bb}) cancel each other and disappear from the resulting
matrix element, so there contributes only the first diagram in the
decomposition in Fig.~\ref{flvertex}.

As a consequence of the gauge invariance the charmonium production
amplitude for on-mass-shell quark and antiquark states obeys the
following property
\begin{eqnarray*}
\bar{u}(k_1)\Psi^{c_2c_1}v(k_2)\rightarrow 0\quad{\rm if}\quad
q_{1,t}\;\mathrm{or}\;q_{2,t}\rightarrow 0.
\end{eqnarray*}

Projection of the hard amplitude onto the singlet charmonium bound
state $V_{\mu\nu}^{c_{1}c_{2}}$ is given by a 4-dimentional
integral over relative momentum of quark and antiquark
$q=(k_1-k_2)/2$ \cite{HKSST00,HKSST01}:
\begin{eqnarray}\nonumber
&&V_{J,\,\mu\nu}^{c_{1}c_{2}}(q_1,q_2)={\cal
P}(q\bar{q}\rightarrow\chi_{cJ})\bullet
\Psi^{c_1c_2}_{ik,\,\mu\nu}(k_1,k_2)=
2\pi\cdot\sum_{i,k}\sum_{L_{z},S_{z}}\frac{1}{\sqrt{m}}\int
\frac{d^{\,4}q}{(2\pi )^{4}}\delta \left(
q^{0}-\frac{{\bf q}^{2}}{M}\right)\times\\
&&\times\,\Phi_{L=1,L_{z}}({\bf q})\cdot\left\langle
L=1,L_{z};S=1,S_{z}|J,J_{z}\right\rangle \left\langle
3i,\bar{3}k|1\right\rangle {\rm
Tr}\left\{\Psi_{ik,\,\mu\nu}^{c_{1}c_{2}}{\cal
P}_{S=1,S_{z}}\right\}, \label{amplitude-diff} \\
&&\Psi_{ik,\,\mu\nu}^{c_{1}c_{2}}=-g^2\biggl[t^{c_1}_{ij}t^{c_2}_{jk}\cdot
\biggl\{\gamma_{\nu}\frac{\hat{q}_{1}-\hat{k}_{1}-m}
{(q_1-k_1)^2-m^2}\gamma_{\mu}\biggr\}-t^{c_2}_{kj}t^{c_1}_{ji}\cdot
\biggl\{\gamma_{\mu}\frac{\hat{q}_{1}-\hat{k}_{2}+m}{(q_1-k_2)^2-m^2}
\gamma_{\nu}\biggr\}\biggr].\nonumber
\end{eqnarray}
Here the function $\Phi_{L=1,L_{z}}({\bf q})$ is the momentum
space wave function of the charmonium, and the projection operator
${\cal P}_{S=1,S_{z}}$ for a small relative momentum $q$ has the
form
\begin{eqnarray}
{\cal
P}_{S=1,S_{z}}=\frac{1}{2m}(\hat{k}_2-m)\frac{\hat{\epsilon}(S_{z})}
{\sqrt{2}}(\hat{k}_1+m) \; .
\end{eqnarray}

Factor $2\pi$ in the formula (\ref{amplitude-diff}) has been
introduced to compensate the difference between the analogous
formula in \cite{Guberina,Cho-Leibo} written in 3-dimentional
form. The Clebsch-Gordan coefficient in color space in our case is
$\left\langle3i,\bar{3}k|1\right\rangle=\delta^{ik}/\sqrt{N_{c}},$
where factor $1/\sqrt{N_{c}}$ provides the averaging of the
production matrix element squared over intermediate quark states
in the loop. Using that we get
\begin{eqnarray}\nonumber
{\rm Tr}(\Psi\,{\cal
P}_{S=1,S_{z}})&=&-\delta^{c_1c_2}\frac{\epsilon^{\rho}(S_{z})}
{\sqrt{2N_c}}\frac{g^2}{4m}\,{\rm Tr}\biggl\{
\biggl(\gamma_{\nu}\frac{\hat{q}_{1}-\hat{k}_{1}-m}
{(q_1-k_1)^2-m^2}\gamma_{\mu}-\gamma_{\mu}\frac{\hat{q}_{1}-
\hat{k}_{2}+m}{(q_1-k_2)^2-m^2}\gamma_{\nu}\biggr)\times\\
&\times&(\hat{k}_2-m)\gamma_{\rho}(\hat{k}_1+m)\biggr\}.
\label{traceR}
\end{eqnarray}
Since $P$-wave function $\Phi_{L=1,L_{z}}$ vanishes at the origin,
we may expand the trace in (\ref{amplitude-diff}) in a Taylor
series around ${\bf q}=0$, and only the linear terms in
$q^{\sigma}$ in the trace (\ref{traceR}) survive. This yields an
expression proportional to
\begin{eqnarray}\label{expansion}
\int \frac{d^{3}{\bf
q}}{(2\pi)^{3}}q^{\sigma}\Phi_{L=1,L_{z}}({\bf q})=-i
\sqrt{\frac{3}{4\pi}}\epsilon^{\sigma}(L_{z}){\cal R}^{\prime}(0),
\end{eqnarray}
with the derivative of the $P$-wave radial wave function at the
origin ${\cal R}^{\prime}(0)$ whose numerical values can be found
in \cite{EQ95}. The general $P$-wave result (\ref{amplitude-diff})
may be further reduced by employing the Clebsch-Gordan identity
which for the scalar charmonium $\chi_{cJ=0}$ reads
\begin{eqnarray*}
&&{\cal
T}^{\sigma\rho}_{J=0}\equiv\sum_{L_{z},S_{z}}\!\!\left\langle
1,L_{z};1,S_{z}|0,0\right\rangle\epsilon^{\sigma}(L_{z})\epsilon^{\rho}(S_{z})
\!=\!\sqrt{\frac13}\biggl(g^{\sigma\rho}-\frac{P^{\,\sigma}P^{\,\rho}}{M^2}\biggr).
\end{eqnarray*}

Using the relations (\ref{dec}) and $q_1-q_2=P_{M}$ and $s\gg
|{\bf q}_{0,t}|^2,$ we obtain
\begin{eqnarray}\label{sx1x2}
(q_1q_2)=\frac12(q_{1,t}^2+q_{2,t}^2-M^2),\quad
s\,x_1x_2=M^2+|{\bf P}_{M,t}|^2\equiv M_{\perp}^2\,,
\end{eqnarray}
that will be useful below. The longitudinal momentum fractions are
now calculated as
\begin{eqnarray}
x_{1,2} &=& \frac{\sqrt{M^2+|{\bf P}_{M,t}|^2}}{\sqrt{s}}
\exp(\pm y)\nonumber\,,\\
x_{1,2}' &=& x_0=\frac{|{\bf q}_{\,0,t}|}{\sqrt{s}}\,.
\label{x1_x2}
\end{eqnarray}
Above $y$ is the rapidity of the produced meson.

Therefore, within QMRK approach we get finally the following
vertex function
\begin{eqnarray}  \nonumber
&&{}V^{c_1c_2}_{J=0}(q_1,q_2)=8ig^2\frac{\delta^{c_1c_2}}{M}
\frac{{\cal R}^{\prime}(0)}{\sqrt{\pi M N_c}}
\frac{3M^2(q_{1,t}q_{2,t})+2q_{1,t}^2q_{2,t}^2-(q_{1,t}q_{2,t})(q_{1,t}^2+q_{2,t}^2)}{(M^2-q_{1,t}^2-q_{2,t}^2)^2}.
\label{major-vert-J0}
\end{eqnarray}

We have also calculated the subprocess matrix element squared
${\cal B}(q_1,q_2)=V_{\mu\nu}V^{\mu\nu}$ that is usually used in
inclusive production calculations. The form of ${\cal B}(q_1,q_2)$
is identical to the form of the matrix element squared obtained
very recently by Likhoded and Luchinsky in Ref.~\cite{Likhoded}.

The objects $f_{g,1}^{off}(x_1,x_1',q_{0,t}^2,q_{1,t}^2,t_1)$ and
$f_{g,2}^{off}(x_2,x_2',q_{0,t}^2,q_{2,t}^2,t_2)$ appearing in
formula (\ref{ampl}) are skewed (or off-diagonal) unintegrated
gluon distributions. They are non-diagonal both in $x$ and $q_t^2$
space. Usual off-diagonal gluon distributions are non-diagonal
only in $x$. In the limit $x_{1,2} \to x_{1,2}'$, $ q_{0,t}^2 \to
q_{1/2,t}^2$ and $t_{1,2} \to 0$ they become usual UGDFs. Our
choice of different UGDFs will be discussed in more detail in a
separate section.

\subsection{Khoze-Martin-Ryskin approach}

In the original Khoze-Martin-Ryskin (KMR) approach \cite{KMR}
the amplitude is written as
\begin{equation}
{\cal
  M}=N\int\frac{d^2q_{0,t}P[\chi_c(0^+)]}{q_{0,t}^2q_{1,t}^2q_{2,t}^2}
f_g^{KMR}(x_1,x'_1,Q_{1,t}^2,\mu^2;t_1)f_g^{KMR}(x_2,x'_2,Q_{2,t}^2,\mu^2;t_2)
\; ,
\label{KMR_amplitude}
\end{equation}
where only one transverse momentum is taken into account somewhat
arbitrarily as
\begin{eqnarray}
Q_{1,t}^2=\mathrm{min}\{q_{0,t}^2,q_{1,t}^2\} \; , \qquad
Q_{2,t}^2=\mathrm{min}\{q_{0,t}^2,q_{2,t}^2\} \; ,
\label{glue_momenta}
\end{eqnarray}
and the normalization factor $N$ can be written in terms
of the $\chi_c(0^+)\to gg$ decay width (see below).

In the KMR approach the large meson mass approximation
$M\gg |{\bf q}_{1,t}|,\,|{\bf q}_{2,t}|$ is adopted, so
the gluon virtualities are neglected in the vertex factor
\begin{equation}
P[\chi_c(0^+)]\simeq(q_{1,t}q_{2,t})=(q_{0,t}+p'_{1,t})(q_{0,t}-p'_{2,t}).
\label{KMR_vert}
\end{equation}
In our approach we generalize the approximation taking into account
corresponding off-shell effects.

The KMR UGDFs are written in the factorized form:
\begin{equation}
f_g^{KMR}(x,x',Q_t^2,\mu^2;t)=f_g^{KMR}(x,x',Q_t^2,\mu^2)\exp(b_0t)
\end{equation}
with $b_0=2$ GeV$^{-2}$ \cite{KMR}. In our approach we use
different parametrization of the $t$-dependent isoscalar form factors
(see Eqns. (\ref{skewed_UGDFs}) and (\ref{off-diag-formfactors}) below).

Please note that the KMR and our skewed UGDFs have different
number of arguments. In the KMR approach there is only one effective
gluon transverse momentum (see Eq.(\ref{glue_momenta})) compared to two
idependent transverse momenta in our case (see Eq.(\ref{skewed_UGDFs})).

The KMR skewed distributions are given in terms of conventional
integrated densities $g$ and the so-called Sudakov form factor $T$ as
follows:
\begin{equation}
f_g^{KMR} (x,x',Q_t^2, \mu^2) = R_g
\frac{\partial}{\partial \ln Q_t^2}
\left[
\sqrt{T(Q_t^2,\mu^2)} x g(x,Q_t^2)
\right] \; .
\label{KMR_UGDF}
\end{equation}
The square root here was taken using arguments that only survival
probability for hard gluons is relevant.
It is not so-obvious if this approximation is reliable for $c \bar c$
quarkonium production.
In addition this has to be contrasted with the choice
of gluon momentum of the KMR UGDF in (\ref{glue_momenta}) as minimal
(not harder) of two gluons.
The factor $R_g$ approximately accounts for the single $\log Q^2$ skewed
effect \cite{KMR}. Please note also that in contrast to our approach
the skewed KMR UGDF does not explicitly depend on $x'$
(assuming $x' \ll x \ll 1$). Usually this factor is estimated to be
1.3--1.5. In our evaluations here we take it to be equal 1 to avoid
further uncertainties.

In contrast to the Higgs case, in the case of light quarkonium production
rather small values of gluon transverse momenta give the dominant
contribution to the integral (\ref{KMR_amplitude}). Therefore it
becomes essential what one does phenomenologically in the
nonperturbative region $Q_t^2 < Q_{0}^2$,
where the $Q_0^2$ is a minimal nonperturbative
scale for standard integrated distributions.
Of course, formally for scales smaller than $Q_0^2$ the standard
collinear gluon distributions do not exist and an extra extension is
unavoidable. This issue was not discussed in detail in the
literature. We shall illustrate this point in the result section.

The Sudakov factor is the result of resumming the virtual contributions
in the DGLAP evolution and reads
\begin{equation}
T(Q_t^2,\mu^2) = \exp \left(
- \int_{Q_t^2}^{\mu^2} \frac{\alpha_s(k_t^2)}{2 \pi} \frac{d
  k_t^2}{k_t^2}
\int_{0}^{1-\Delta} [ z P_{gg}(z) + \sum_q P_{qg}(z) ] dz
\right) \; ,
\label{Sudakov_ff}
\end{equation}
with $\Delta = k_t/(\mu + k_t)$. In their (KMR) estimates the hard
scale is usually taken $\mu = M_{\chi}/2$. Of course the choice of the
scale is somewhat arbitrary, and the consequences of this choice were
not discussed in the literature.

\begin{figure}[h!]    
\includegraphics[width=0.4\textwidth]{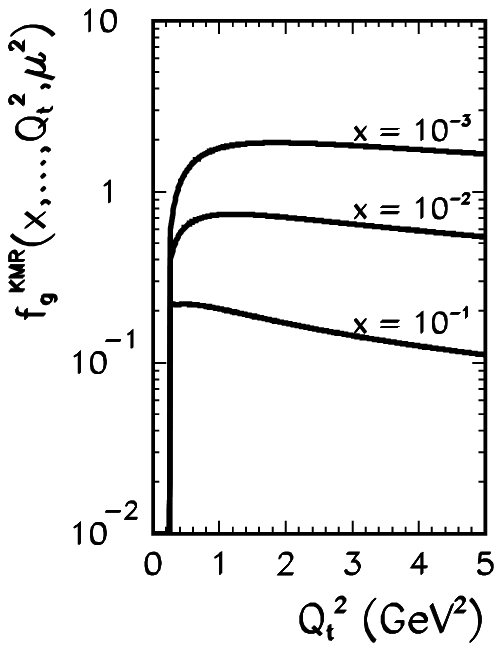}
\includegraphics[width=0.4\textwidth]{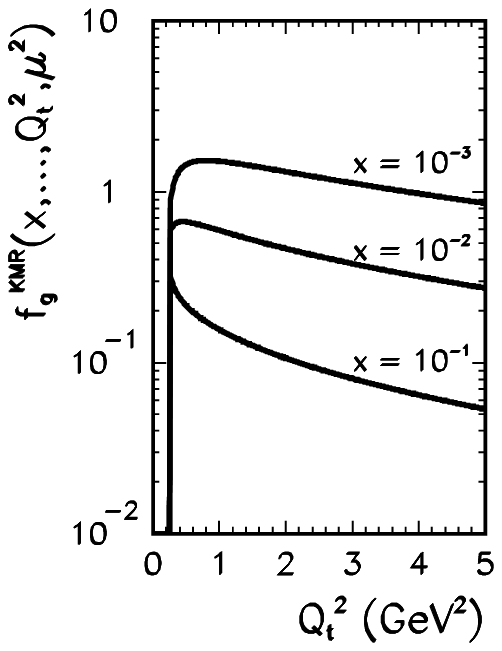}
\caption{KMR distribution as a function of effective transferse momentum
$Q_t^2$ for different values of $x$ and $\mu^2 = M_{\chi}^2$ (left panel)
and $\mu^2 = (M_{\chi}/2)^2$ (right panel).
}
\label{fig:f_kmr}
\end{figure}

In Fig.\ref{fig:f_kmr} we show the KMR distribution (\ref{KMR_UGDF})
as a function of $Q_t^2$ for different values of $x$ specified
in the figure for two different choices of the scale. The DGLAP distribution
for $Q_t^2 < Q_0^2$ is not well defined and we arbitrarily put it to
zero as in Ref.\cite{KMR-chi}. In principle, one could try several other
extrapolations into the nonperturbative region to see its influence on
the resulting differential cross sections.

\subsection{Our vertex versus Khoze-Martin-Ryskin
vertex}

We wish to compare our vertex function (\ref{major-vert-J0}) with
the KMR vertex function \cite{KMR-chi}. In the KMR limit of large
meson mass $M\gg |{\bf q}_{1,t}|,\,|{\bf q}_{2,t}|$ and
$s\,x_1x_2\simeq M^2$ the vertex reads
\begin{eqnarray}  \nonumber
V^{c_1c_2}_{J=0}[M\gg q_{1,t},\,q_{2,t}]&\simeq&
8ig^2\delta^{c_1c_2}\frac{{\cal R}^{\prime}(0)}{M^3}
\frac{1}{\sqrt{\pi M N_c}} \biggl\{3(q_{1,t}q_{2,t})\biggr\}=\\
&=&i\delta^{c_1c_2}\cdot 8g^2\sqrt{\frac{3}{\pi M}}\frac{{\cal
R}'(0)}{M^3}\cdot (q_{1,t}q_{2,t}). \label{major-vert-J0-KMR}
\end{eqnarray}
Following now the KMR notations \cite{KMR-chi} we write the total
NRQCD QMRK amplitude (\ref{ampl}) (averaged over colour and
polarisation states of incoming gluons) in the limit $M\gg
q_{1,t},\,q_{2,t}$ as
\begin{eqnarray}
{\cal M}=A\,\pi^2\,\frac{s}{2}\,\int d^2
q_{0,t}P[\chi_c(0^+)]\frac{f^{off}_{g,1}(x_1,x_1',q_{0,t}^2,
q_{1,t}^2,t_1)f^{off}_{g,2}(x_2,x_2',q_{0,t}^2,q_{2,t}^2,t_2)}
{q_{0,t}^2\,q_{1,t}^2\, q_{2,t}^2}\,, \label{ampl-KMR}
\end{eqnarray}
where the normalization is
\begin{eqnarray}
A=4g^2\sqrt{\frac{3}{\pi M}}\frac{{\cal R}'(0)}{M^3}\,,
\label{norm-NR}
\end{eqnarray}
and the vertex factor is defined in (\ref{KMR_vert}).

Normalization constant $A$ can be obtained in another way in terms
of the partial decay width $\Gamma(\chi_{c0}\rightarrow gg)$ as
(see formula (19) in the KMR paper \cite{KMR-chi})
\begin{eqnarray}
A^2=K\frac{64\pi\Gamma(\chi_{c0}\rightarrow
gg)}{(N_c^2-1)M^3},\qquad \textrm{NLO}\quad\rightarrow\quad
K=1.5\, . \label{A-from-Gamma}
\end{eqnarray}
Using the expression for $\Gamma(\chi_{c0}\rightarrow gg)$,
obtained in the framework of pNRQCD \cite{Cakir,Close}
\begin{eqnarray}
&&\Gamma(\chi_{c0}\to gg) = 32N_c\alpha_{s}^2\frac{|{\cal
R}'(0)|^2}{M^4},\qquad N_c=3
\end{eqnarray}
we get the normalization constant (with $K=1$)
\begin{eqnarray}
A=4g^2\sqrt{\frac{3}{\pi M}}\frac{{\cal R}'(0)}{M^3}
\end{eqnarray}
which coincides with the normalisation of the vertex factor
obtained within the QMRK approach (\ref{norm-NR}).

Therefore, in the leading order the QMRK approach is in agreement
with the KMR approach in the limit of large meson mass $M\gg |{\bf
q}_{1,t}|,\,|{\bf q}_{2,t}|$. We shall discuss deviations from this
approximation due to the off-shell effects in the result section.
Similar analysis of off-shell effects was performed recently
for inclusive Higgs production in Ref.~\cite{PTS06}.

\subsection{Off-diagonal UGDFs and choice of scales}

In the present work we shall use a few sets of unintegrated gluon
distributions which aim at the description of phenomena where
small gluon transverse momenta are involved. Some details
concerning the distributions can be found in Ref.~\cite{LS06}. We
shall follow the notation there.

The larger energies, the smaller values of parton momentum
fractions come into game. Therefore at larger energies we shall
use distributions constructed exclusively for small values of $x$.
Two of them are based on the idea of gluon saturation. One of them
was obtained based on a saturation-inspired parametrization of the
dipole-nucleon cross section which leads to a good description of
the HERA data \cite{GBW_glue}. The second one \cite{KL01} was
constructed to describe the inclusive RHIC pion spectra. The third
one is the asymptotic BFKL distribution \cite{BFKL}. We do not
wish to repeat more details here. It can be found in individual
references as well as in Ref.~\cite{LS06} where applications of
UGDFs to $c \bar c$ correlations was discussed.

Due to its simplicity the Gaussian smearing of initial transverse
momenta is a good reference point for other approaches. It allows
to study phenomenologically the role of transverse momenta in
several high-energy processes. We define simple unintegrated gluon
distribution
\begin{eqnarray}
{\cal F}_{g}^{Gauss}(x,k_t^2,\mu_F^2)=xg^{coll}(x,\mu_F^2) \cdot
f_{Gauss}(k_t^2)\;,
\label{Gaussian_UGDF}
\end{eqnarray}
where $g^{coll}(x,\mu_F^2)$ are standard collinear (integrated)
gluon distribution and $f_{Gauss}(k_t^2)$ is a Gaussian
two-dimensional function
\begin{eqnarray}
\begin{split}
f_{Gauss}(k_t^2)=\frac{1}{2\pi\sigma_0^2} \exp\left(-k_t^2/2
\sigma_0^2\right)/\pi\,. \label{Gaussian}
\end{split}
\end{eqnarray}
The UGDF defined by Eq.~(\ref{Gaussian_UGDF}) and (\ref{Gaussian})
are normalized such that
\begin{eqnarray}
\int {\cal F}_{g}^{Gauss}(x,k_t^2,\mu_F^2) \; d k_t^2 = x
g^{coll}(x,\mu_F^2)\;. \label{Gaussian_normalization}
\end{eqnarray}

The UGDFs have the following property
\begin{eqnarray}
f(x,k_t^2) \to 0\, ,
\end{eqnarray}
if $k_t^2 \to$ 0. The small-$k_t^2$ region is of nonperturbative
nature and is rather modelled than derived from pQCD.

The two-scale off-diagonal distributions require a separate
discussion. In the general case we do not know UGDFs very well. It
seems reasonable, at least in the first approximation, to take in
the amplitude (\ref{ampl})
\begin{eqnarray}\nonumber
f_{g,1}^{off} &=& \sqrt{f_{g}^{(1)}(x_1',q_{0,t}^2,\mu_0^2) \cdot
f_{g}^{(1)}(x_1,q_{1,t}^2,\mu^2)} \cdot F_1(t_1)\,, \\
f_{g,2}^{off} &=& \sqrt{f_{g}^{(2)}(x_2',q_{0,t}^2,\mu_0^2) \cdot
f_{g}^{(2)}(x_2,q_{2,t}^2,\mu^2)} \cdot F_1(t_2)\,,
\label{skewed_UGDFs}
\end{eqnarray}
where $F_1(t_1)$ and $F_1(t_2)$ are isoscalar nucleon form factors
\cite{DL}
\begin{eqnarray}
F_1(t_{1,2}) = \frac{4 m_p^2 - 2.79\,t_{1,2}} {(4 m_p^2
-t_{1,2})(1-t_{1,2}/071)^2} \;, \label{off-diag-formfactors}
\end{eqnarray}
and $t_1$ and $t_2$ are total four-momentum transfers in the first
and second proton line, respectively. The proton form factor in
the form (\ref{off-diag-formfactors}) gives rather good description of
the $t$-dependence of the elastic $pp$ cross section at high
energies, i.e. for kinematics similar as in our case. The above
prescription for UGDFs (\ref{skewed_UGDFs}) is a bit arbitrary,
although it is inspired by the positivity constraints for {\it
collinear} Generalized Parton Distributions \cite{posit}. It
provides, however, an interpolation between different $x$ and
$q_t^2$ values. Our prescription is more symmetric in variables of
the first and second exchange than the one used in
Ref.~\cite{BBKM06} for Higgs boson production.

The choice of the (factorisation) scales here is not completely
obvious too. We shall try the following three choices:
\begin{eqnarray}
&(1)&\; \mu_0^2 = M^2,\qquad \mu^2 = M^2 , \nonumber \\
&(2)&\; \mu_0^2 = Q_0^2,\qquad \mu^2 = M^2 ,\nonumber  \\
&(3)&\; \mu_0^2 = q_{0,t}^2\;(\mathrm{+freezing\, at}\; q_{0,t}^2
< Q_0^2),\quad  \mu^2 = M^2 \; .
\label{Gaussian_scales}
\end{eqnarray}

The first choice is similar as in Ref.~\cite{KMR,KimbMR}. However,
it is not obvious if the scale associated with the ``hard''
production ($g^* g^* \to \chi_c$) can be used for the left part of
the gluonic ladder where no obvious hard scale appears. Therefore
we shall try also the second choice where we shall use $Q_0^2$ =
0.26 GeV$^2$, i.e. the nonperturbative input for the QCD evolution
in Ref.~\cite{GRV}. Another option was proposed by Lonnblad and
Sj\"odahl in Ref.~\cite{Lonnblad}. They take $q_{0,t}^2$ as a
first scale. In our case this prescription must be supplemented by
freezing the scale for gluon transverse momenta smaller than $Q_0$
(minimal perturbative scale).

When inspecting Eqs.~(\ref{ampl}) and (\ref{skewed_UGDFs}) it
becomes clear that the cross section for elastic
double-diffractive production of a meson is much
more sensitive to the choice of UGDFs than the inclusive cross
sections.

\section{$\gamma^* \gamma^*$ fusion mechanism}

\begin{figure}[h!]    
\centerline{\epsfig{file=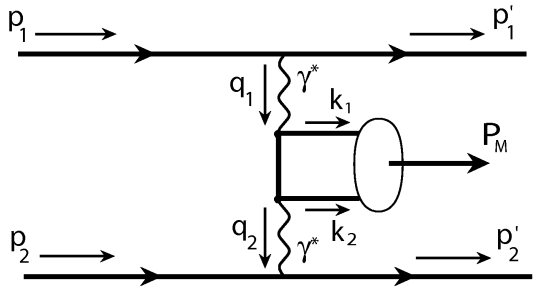,width=7cm}}
\caption{Kinematics of exclusive $\gamma^*\gamma^*$ fusion
mechanism of $\chi_c$-meson production.}
\label{fig:kinematics_gamgam}
\end{figure}

As stated in the introduction we wish to investigate the
competition of the diffractive QCD mechanism discussed in the
previous sections and the $\gamma^*\gamma^*$-fusion mechanism
shown in more detail in Fig.~\ref{fig:kinematics_gamgam}.

\subsection{NRQCD approach}

In the most general case the Born amplitude reads
\begin{eqnarray}
{\cal M}_{\lambda_1,\lambda_2,\lambda_1',\lambda_2'} &=& \pm e^2\{
\overline{u}(p'_1,\lambda_1') [F_1(t_1) \gamma^{\nu}  \pm
i\frac{\sigma^{\nu \nu''}}{2 m_N} q_{1,\nu''} F_2(t_1) ]
u(p_1,\lambda_1) \}\; \frac{g_{\nu \nu'}}{t_1}\;
V_{\mu'\nu'}^{\gamma^* \gamma^* \to \chi_{cJ}}(k_1,k_2)\;
\frac{g_{\mu \mu'}}{t_2} \nonumber\\
&&{}\{ \overline{u}(p'_2,\lambda_2') [F_1(t_2) \gamma^{\mu} \pm
i\frac{\sigma^{\mu \mu''}}{2 m_N} q_{1,\mu''} F_2(t_2) ]
u(p_2,\lambda_2) \}\,, \label{QED_full_amplitude}
\end{eqnarray}
where the sign $``+"$ stands for $pp$-scattering and the sign
$``-"$ for $p\bar{p}$-scattering. In the following we have omitted
the spin-flipping contributions related to the respective Pauli
form factors. Limiting to large energies ($\sqrt{s} \gg M + 2m_N$)
and small momenta transfer $t_{1,2}$ ($|t_{1,2}| \ll 4 m_N^2$) the
matrix element for $pp \to pp\,\chi_{cJ}$ reaction via
$\gamma^*\gamma^*$-fusion can be written as
\begin{eqnarray}
{\cal M}^{\gamma^* \gamma^*} \approx e F_1(t_1) \frac{(p_1 +
p_1')^{\nu}}{t_1}\, V_{\mu\nu}^{\gamma^* \gamma^* \to\,
\chi_{cJ}}(q_1,q_2)\, \frac{(p_2 + p_2')^{\mu}}{t_2}\, e
F_1(t_2)\,, \label{gamma_gamma_eta_amplitude}
\end{eqnarray}
where $F_1(t_1)$ and $F_1(t_2)$ are Dirac proton electromagnetic
form factors, and the $\gamma^*\gamma^* \to \chi_{cJ}$ vertex has
analogous form as (\ref{amplitude-diff})
\begin{eqnarray}\nonumber
&&V_{\mu\nu}^{\gamma^* \gamma^* \to\, \chi_{cJ}}(q_1,q_2)={\cal
P}(q\bar{q}\rightarrow\chi_{cJ})\bullet
\Psi_{\mu\nu}(k_1,k_2)=2\pi\cdot\sum_{i,\,k}\sum_{L_{z},S_{z}}\frac{1}{\sqrt{m}}\int
\frac{d^{\,4}q}{(2\pi )^{4}}\delta \left(
q^{0}-\frac{{\bf q}^{2}}{M}\right)\times\\
&&\times\,\Phi_{L=1,L_{z}}({\bf q})\cdot\left\langle
L=1,L_{z};S=1,S_{z}|J,J_{z}\right\rangle \left\langle
3i,\bar{3}k|1\right\rangle {\rm Tr}\left\{\Psi_{\mu\nu}^{ik}{\cal
P}_{S=1,S_{z}}\right\}, \label{amplitude-gamgam} \\
&&\Psi_{\mu\nu}^{ik}=\delta^{ik}
\left(\frac{2e}{3}\right)^2\biggl[\gamma_{\nu}\frac{\hat{q}_{1}-\hat{k}_{1}-m}
{(q_1-k_1)^2-m^2}\gamma_{\mu}-\gamma_{\mu}\frac{\hat{q}_{1}-
\hat{k}_{2}+m}{(q_1-k_2)^2-m^2}\gamma_{\nu}\biggr],\quad
\left\langle 3i,\bar{3}k|1\right\rangle=
\frac{\delta^{ik}}{\sqrt{N_c}},\nonumber
\end{eqnarray}
and $\delta^{ik}\delta_{ik}=N_c.$ Repeating similar steps as for
diffractive production we get
\begin{eqnarray}\nonumber
&&{}V_{\mu\nu}^{\gamma^* \gamma^* \to\,
\chi_{cJ}}(q_1,q_2)=-2i\left(\frac{2e}{3}\right)^2\frac{{\cal
R}^{\prime}(0)}{M}\sqrt{\frac{3N_c}{\pi M}}{\cal
T}^{\sigma\rho}_{J} \biggl\{\frac{1}{(q_1q_2)}
\biggl[2(q_{2,\nu}-q_{1,\nu})q_{1,\rho}g_{\mu\sigma}-\\
&&{}-(q_{1,\mu}-q_{2,\mu})(q_{2,\sigma}+q_{1,\sigma})g_{\nu\rho}+
\dfrac12(q_{1,\rho}-q_{2,\rho})(q_{1,\sigma}-q_{2,\sigma})g_{\mu\nu}+
(M^2+q_1^2-q_2^2)g_{\nu\rho}g_{\mu\sigma}+\nonumber\\
&&{}+\{q_1,\,\mu\leftrightarrow q_2,\,\nu\}\biggr]+
\frac{q_{1,\sigma}+q_{2,\sigma}}{2(q_1q_2)^2}\biggl[(q_{1,\rho}-q_{2,\rho})
(q_{1,\mu}q_{1,\nu}-q_{2,\mu}q_{2,\nu})+\\
&&{}+(M^2-q_1^2+q_2^2)q_{1,\rho}g_{\mu\nu}-2M^2q_{1,\mu}g_{\nu\rho}+
\{q_1,\,\mu\leftrightarrow q_2,\,\nu\}\biggr]\biggr\} \; .\nonumber
\end{eqnarray}
For example, in the case of scalar meson $(J=0)$ we have
\begin{eqnarray}
\label{major-vert-gamgam} &&{}V_{\mu\nu}^{\gamma^* \gamma^* \to\,
\chi_{c0}}(q_1,q_2)=-4i\left(\frac{2e}{3}\right)^2\frac{{\cal
R}^{\prime}(0)}{M}\sqrt{\frac{N_c}{\pi
M}}\biggl\{\frac{M^2g_{\mu\nu}-(q_{1,\mu}-q_{2,\mu})(q_{1,\nu}-q_{2,\nu})}
{(q_1q_2)}-\nonumber\\
&&{}-\frac{1}{4(q_1q_2)^2}\biggl[[M^2(M^2-2q_2^2-2q_1^2)+(q_1^2-q_2^2)^2]g_{\mu\nu}+
2(M^2+q_1^2-q_2^2)q_{2,\mu}q_{2,\nu}+\nonumber\\
&&{}+2(M^2+q_2^2-q_1^2)q_{1,\nu}q_{1,\mu}+4M^2q_{2,\nu}q_{1,\mu}\biggr]
\biggr\}.
\end{eqnarray}
This amplitude can be rewritten in the general gauge invariant
form \cite{PTS06} in terms of two independent form factors
\begin{eqnarray}\nonumber
&&{}V_{\mu\nu}^{\gamma^* \gamma^* \to\,
\chi_{c0}}(q_1,q_2)=-4i\left(\frac{2e}{3}\right)^2\frac{{\cal
R}^{\prime}(0)}{M}\sqrt{\frac{N_c}{\pi M}}\biggl\{F_1(q_1,q_2)
((q_1q_2)g_{\mu\nu}-q_{1,\mu}q_{2,\nu})+\\
&&{}+F_2(q_1,q_2)
\left(q_{1,\nu}q_{2,\mu}-\frac{q_1^{2}}{(q_1q_2)}\,q_{2,\mu}q_{2,\nu}-\frac{q_2^{2}}{(q_1q_2)}\,q_{1,\mu}q_{1,\nu}+
\frac{q_1^{2}q_2^{2}}{(q_1q_2)^2}\,q_{1,\mu}q_{2,\nu}\right)\biggr\},
\label{gener}
\end{eqnarray}
where
\begin{eqnarray*}
F_1(q_1,q_2)=\frac{q_1^2q_2^2+(q_1q_2)(q_1^2+q_2^2-3(q_1q_2))}{(q_1q_2)^3},\quad
F_2(q_1,q_2)=\frac{1}{(q_1q_2)}.
\end{eqnarray*}
From this amplitude the standard decay width follows
\begin{eqnarray}
\Gamma(\chi_{c0}\rightarrow
\gamma\gamma)=\frac{256}{3}\,\alpha_{em}^2\frac{|{\cal
R}'(0)|^2}{M^4},
\end{eqnarray}
So the normalization of the amplitude (\ref{gener}) is correct.

Analogously with the diffractive case (\ref{dec}) let us define
the photon transverse momenta. Momentum conservation dictates us
the following decompositions of photon momenta into longitudinal
and transverse parts
\begin{eqnarray}
q_1=x_1p_1+\frac{t_1}{s}\,p_2+q_{1,t},\quad
q_2=-x_2p_2-\frac{t_2}{s}\,p_1+q_{2,t},\quad
q_{1/2,t}^2 \simeq t_{1,2}(1-x_{1,2}) \; ,
\end{eqnarray}
where $t_{1,2}\equiv q_{1,2}^2$. Due to the gauge invariance
we have similarly to (\ref{decomp})
\begin{eqnarray}
(p_1+p_1')^{\nu}V_{\mu\nu}(p_2+p_2')^{\mu}=4p_1^{\nu}p_2^{\mu}V_{\mu\nu}.
\end{eqnarray}
In the relevant limit $t_{1,2}\ll x_{1,2}s$ we get finally the
following matrix element for $pp \to pp\,\chi_{c}(0)$ reaction via
$\gamma^*\gamma^*$-fusion
\begin{eqnarray}
{\cal M}^{\gamma^* \gamma^*}&\approx&
-i\,4s\left(\frac{4e^2}{3}\right)^2\frac{{\cal
R}^{\prime}(0)}{M}\sqrt{\frac{N_c}{\pi M}}\frac{F_1(t_1)}{t_1}
\frac{F_1(t_2)}{t_2}
\frac{3M^2(q_{1,t}q_{2,t})+2t_1t_2-(q_{1,t}q_{2,t})(t_1+t_2)}{(M^2-t_1-t_2)^2}
\; ,
\label{gamma-gamma-chic-amp}
\end{eqnarray}
where $(q_{1,t}q_{2,t})=-\sqrt{t_1t_2(1-x_1)(1-x_2)}\cos{\Phi},$
and $\Phi$ is the relative angle between photons (or outgoing
protons). The amplitude (\ref{gamma-gamma-chic-amp}) is purely
imaginary, so there is no interference with diffractive process
describing by purely real amplitude (\ref{ampl}). This amplitude will
be used in the following to calculate differential cross section.

\subsection{Equivalent Photon Approximation}

In the equivalent photon approximation (EPA) the total cross section
for $p p \to p \chi_c(0) p$ can be written as a convolution
of EPA flux factors and the $\gamma \gamma \to \chi_c(0)$ resonant
cross section
\begin{equation}
\sigma = \int dz_1 dz_2
\left(
\frac{dn}{dz_1}(z_1)
\frac{dn}{dz_2}(z_2)
\sigma(\gamma \gamma \to \chi_c(0))
\right)
\; .
\label{sigma_tot_EPA}
\end{equation}
The elementary cross section can be written in terms of partial
decay width as
\begin{equation}
\sigma(\gamma \gamma \to \chi_c(0)) \approx \frac{4 \pi^2}{M_R^2}
\Gamma_{\chi_c(0) \to \gamma \gamma} \delta(M - M_R) \; .
\label{sigma_ele}
\end{equation}
Let us introduce two new variables:
\begin{equation}
\begin{split}
x_F &= z_1 - z_2 \; \\
M &= \sqrt{s z_1 z_2} \; .
\end{split}
\label{new_variables}
\end{equation}
Now the cross section can be written as
\begin{equation}
\frac{d \sigma}{d x_F d M} =
\frac{2 M}{s (z_1+z_2)} \;
\frac{dn}{dz_1}(z_1) \;
\frac{dn}{dz_2}(z_2) \;
\frac{4 \pi^2}{M_R^2}
\Gamma_{\chi_c(0) \to \gamma \gamma} \delta(M - M_R) \; .
\end{equation}
Integrating over invariant mass of the two photons we get
\begin{equation}
\frac{d \sigma}{d x_F} =
\frac{2 M_R}{s (z_1+z_2)} \;
\frac{dn}{dz_1}(z_1) \;
\frac{dn}{dz_2}(z_2) \;
\frac{4 \pi^2}{M_R^2}
\Gamma_{\chi_c(0) \to \gamma \gamma} \; ,
\end{equation}
where now
\begin{equation}
\begin{split}
z_1 = \frac{1}{2}x_F + \frac{1}{2}\sqrt{x_F^2 + 4 M_R^2/s} \; , \\
z_2 = -\frac{1}{2}x_F + \frac{1}{2}\sqrt{x_F^2 + 4 M_R^2/s} \; .
\end{split}
\end{equation}
This equation is suitable to calculate distribution of $\chi_c(0)$
in the Feynman variable $x_F$. The analytical flux factors of photons in
protons of Drees and Zeppenfeld \cite{DZ89} are taken.
We have also tried:
\begin{equation}
f(z) \equiv \frac{dn}{dz}(z) = \int d^2 q_t \;
\frac{q_t^2} {(q_t^2+z^2 m_N^2)^2} \; F_1^2(t) \; ,
\end{equation}
where $t = -(q_t^2 + z m_N^2)/(1-z)$.
The decay width from PDG \cite{PDG06} is
$\Gamma_{\chi_c(0) \to \gamma \gamma}$ = 0.2626 10$^{-5}$ GeV.

\section{ Pomeron-Pomeron fusion}

\begin{figure}[h!]    
\centerline{\epsfig{file=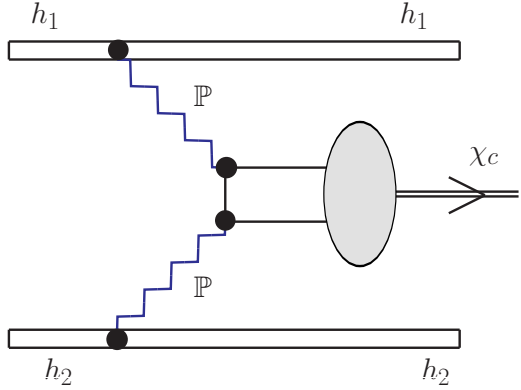,width=7cm}}
\caption{Pomeron-pomeron fusion
mechanism of $\chi_c$-meson production.}
\label{fig:pomeron_pomeron_fusion}
\end{figure}

Above we have shown how to calculate the diffractive $\chi_c(0^+)$ meson
production mechanism in a QCD-inspired approach. Often in the literature
in order to describe the high-energy processes one uses
a phenomenological object known as pomeron. Often a vector nature
is prescribed to such an object, i.e. it is assumed that it couples to
the nucleons or (quarks) via $\gamma_{\mu}$ matrices, i.e. similarly
as photon. The corresponding mechanism for exclusive $\chi_c$ meson production
is sketched in Fig.\ref{fig:pomeron_pomeron_fusion}.

The amplitude for our exclusive process may be written as
\begin{equation}
{\cal M}_{pp\, \to pp\,\chi_{cJ}}^{\mathbb{P} \mathbb{P} \to\,
\chi_{cJ}} \approx A_R(s_1,t_1) (p_1 + p_1')^{\nu} \,
V_{\mu\nu}^{\mathbb{P} \mathbb{P}\to\, \chi_{cJ}}(k_1,k_2)\, (p_2
+ p_2')^{\mu} A_R(s_2,t_2) \; .
\label{pomeron_pomeron_chic0_amplitude}
\end{equation}
In the equation above
$A_R(s_{1/2},t_{1/2})$ are so-called Regge propagators.
They can be written as:
\begin{equation}
A_R(s_{1/2},t_{1/2}) = r_c \frac{C_{\mathbb{P}}}{3}
\left( \frac{s_{1/2}}{s_0} \right)^{\delta} F_1^{\mathbb{P}}(t_{1/2}) \; .
\label{Regge_propagators}
\end{equation}
$F_1^{\mathbb{P}}(t_{1/2})$ are form factors describing
helicity-preserving coupling of the pomeron to the nucleon.
We take them to be identical to the Dirac electromagnetic form factors
of the proton. Such a choice is justified by the phenomenology of
elastic proton-proton scattering (see e.g.\cite{DL}).
The pomeron coupling to the charm quark and/or antiquark is reduced
compared to the coupling to the nucleon ($C_{\mathbb{P}}$) by the factor 3
(three quarks in the nucleon versus single charm quark/antiquark)
and extra factor $r_c$
(heavy quark interaction is weaker than light quark interaction).
This parameter can be extracted from the inelastic interaction of
$J/\psi$ with the nucleons in nuclei \cite{KMR-chi}. In the Tevatron energy
$r_c \sim$ 0.2. The parameters of the pomeron exchanges: $C_{\mathbb{P}}$ and
$\delta$ are taken from the Donnachie-Landshoff fits to the
proton-proton and proton-antiproton total cross sections \cite{DL92}.
The vertex $V^{\mathbb{P}\mathbb{P} \to \chi_c}$ can be obtained from the
vertex $V^{\gamma^* \gamma^* \to \chi_c}$ by replacing charm quark
charge by unity.

The comparison of the results with pomeron-pomeron fusion
with the diffractive QCD results will be given in the
result section.

\section{Cross section and phase space}

The cross section for the 3-body reaction $pp\to pp\chi_{c}$ can
be written as
\begin{eqnarray}
d\sigma_{pp\to pp\chi_{c}}=\frac{1}{2s}\,|{\cal M}|^2\cdot
d^{\,3}PS\,. \label{cross sect}
\end{eqnarray}

The three-body phase space volume element reads
\begin{eqnarray}
d^3 PS = \frac{d^3 p_1'}{2 E_1' (2 \pi)^3} \frac{d^3 p_2'}{2 E_2'
(2 \pi)^3} \frac{d^3 P_M}{2 E_M (2 \pi)^3} \cdot (2 \pi)^4
\delta^4 (p_1 + p_2 - p_1' - p_2' - P_M) \; . \label{dPS_element}
\end{eqnarray}
At high energies and small momentum transfers the phase space
volume element can be written as
\begin{eqnarray}
d^3 PS = \frac{1}{2^8 \pi^4} dt_1 dt_2 d\xi_1 d\xi_2 d \Phi \;
\delta \left( s(1-\xi_1)(1-\xi_2)-M^2 \right) \; ,
\label{dPS_element_he1}
\end{eqnarray}
where $\xi_1$, $\xi_2$ are longitudinal momentum fractions carried
by outgoing protons with respect to their parent protons and the
relative angle between outgoing protons $\Phi \in (0, 2\pi)$.
Changing variables  $(\xi_1, \xi_2) \to (x_F, M^2)$ one gets
\begin{eqnarray}
d^3 PS = \frac{1}{2^8 \pi^4} dt_1 dt_2 \frac{dx_F}{s \sqrt{x_F^2 +
4 (M^2+|{\bf P}_{M,t}|^2)/s}} \; d \Phi \; .
\label{dPS_element_he2}
\end{eqnarray}

It is more convenient for lower (but still high) energy to use
variable $x_F$. However, at very high energies the cross section
becomes too much peaked at $x_F \approx$ 0 due to the jacobian
\begin{eqnarray}
J\approx\frac{1}{\sqrt{x_F^2+4M^2/s}} \to \frac{\sqrt{s}}{2M}
\label{jacobian}
\end{eqnarray}
and the use of rapidity $y$ instead of $x_F$ is recommended. The
phase space element in this case has the following simple form
\begin{eqnarray}
d^3 PS = \frac{1}{2^8 \pi^4\,s} dt_1 dt_2 dy d \Phi\,.
\label{dPS_element_he3}
\end{eqnarray}
If $x_F$ is used then
\begin{eqnarray}
\xi_{1,2} \approx 1 - \frac{1}{2} \sqrt{x_F^2 + \frac{4 M^2}{s}}
\mp \frac{x_F}{2} \; .
\end{eqnarray}
In the other case when the meson rapidity is used then
\begin{eqnarray}
\xi_{1,2} \approx 1 - \frac{M}{\sqrt{s}} \exp(\pm y) \; .
\end{eqnarray}

Now the four-momentum transfers in both proton lines can be
calculated as
\begin{eqnarray}
t_{1,2} = -\frac{p'^2_{1/2,t}}{\xi_{1,2}} - \frac{(1-\xi_{1,2})^2
m_p^2}{\xi_{1,2}} \; . \label{t12_transverse_momenta}
\end{eqnarray}
Only if $\xi_{1,2}$ = 1, $t_{1,2} = -p'^2_{1/2,t}$. The latter
approximate relation was often used in earlier works on
diffractive production of particles. However, in practice
$\xi_{1,2} \ne$ 0 and the more exact equation must be used. The
range of $t_1$ and $t_2$ is not unlimited as it is often assumed.
One can read off from Eq.(\ref{t12_transverse_momenta}) a
kinematical upper limit for $t_{1,2}$ which is
\begin{eqnarray}
t_{1,2} < - \frac{(1-\xi_{1,2})^2}{\xi_{1,2}} m_p^2  \; .
\label{upper_limit_for_t12}
\end{eqnarray}
In practice these phase space limits become active only for
$|x_F|>$ 0.2. The lower limits are energy dependent but are not
active in practice.


\section{Results}

In Ref.\cite{KMR-chi} estimates of the integrated cross sections were
given. We wish to concentrate on differential distributions.
Before we show our results we wish to discuss uncertainties
related to the KMR approach.

\subsection{Uncertainties in the KMR approach}

\begin{figure}[!h]    %
\includegraphics[width=0.55\textwidth]{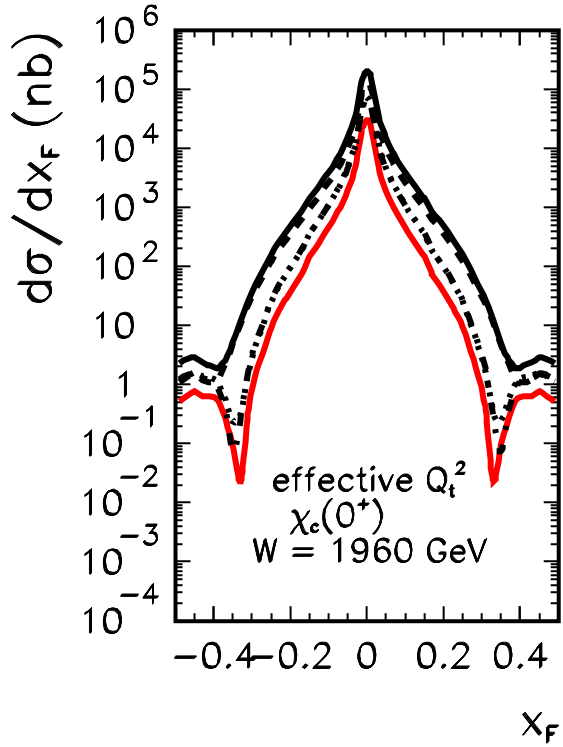}
   \caption{\label{fig:dsig_dxf_kmr_kt2}
   \small  Effect related to the choice of effective transverse
momentum in the KMR UGDF on distribution in Feynman variable $x_F$ in the
KMR approach.
The solid lines are for the choice 1 (upper) and choice 2 (lower)
The dashed line is for choice 3, the dotted line for choice 4 and the
dash-dotted line for the last possibility.
The calculation was done for the Tevatron energy W = 1960 GeV.}
\end{figure}

In the KMR approach only one effective transverse momentum is taken
explicitly in their skewed unintegrated distributions. In the KMR
prescription it is the minimum of the transverse momenta of the two
gluons connected to the same proton line.
In order to see the uncertainties related to such a choice we shall
present results obtained with the following, equally arbitrary, choices:
\begin{itemize}
\item  1) \; $Q_{1,t}^2 = \min(q_{0,t}^2,q_{1,t}^2), \;\; Q_{2,t}^2 =
  \min(q_{0,t}^2,q_{2,t}^2)$ \; ,
\item  2) \; $Q_{1,t}^2 = \max(q_{0,t}^2,q_{1,t}^2), \;\; Q_{2,t}^2 =
  \max(q_{0,t}^2,q_{2,t}^2)$ \; ,
\item  3) \; $Q_{1,t}^2 = q_{1,t}^2, \;\; Q_{2,t}^2 = q_{2,t}^2$ \; .
\item  4) \; $Q_{1,t}^2 = q_{0,t}^2, \;\; Q_{2,t}^2 = q_{0,t}^2$ \; ,
\item  5) \; $Q_{1,t}^2 = (q_{0,t}^2+q_{1,t}^2)/2, \;\; Q_{2,t}^2 =
  (q_{0,t}^2+q_{2,t}^2)/2$ \; .
\end{itemize}
As an example in Fig.\ref{fig:dsig_dxf_kmr_kt2} we show
results for the Feynman $x_F$ distribution. As can be seen from the figure
there are rather large uncertainties related to the choice of the effective
transverse momentum.


\begin{figure}[!h]    %
\includegraphics[width=0.55\textwidth]{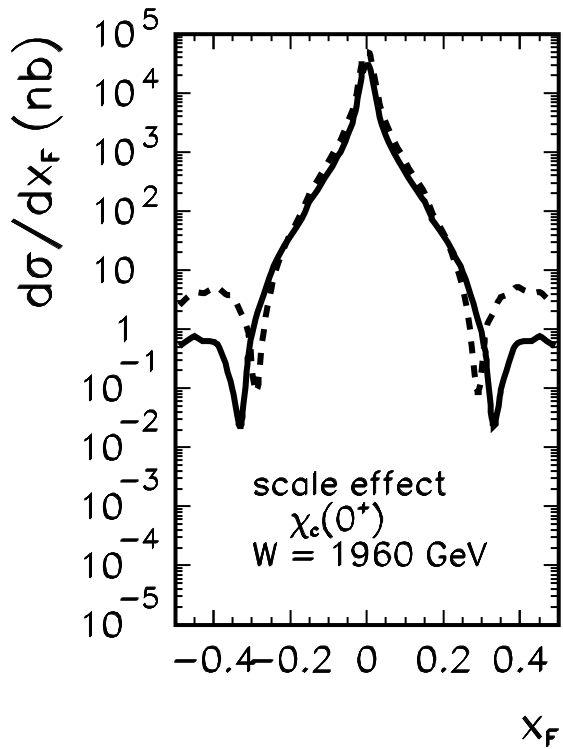}
   \caption{\label{fig:dsig_dxf_kmr_scales}
   \small  Scale effect on distribution in Feynman $x_F$ in the KMR approach
(KMR UGDF, KMR vertex). The solid line is for $\mu^2 = M_{\chi}^2$
and the dashed line is for $\mu^2 = M_{\chi}^2/4$.
The calculation was done for the Tevatron energy W = 1960 GeV.}
\end{figure}

The estimate by Khoze, Martin, Ryskin and Stirling in Ref.\cite{KMR-chi}
was done for one selected value of the hard scale in the KMR UGDF
$\mu^2 = M_{\chi_c(0)}^2/4$. In Fig.\ref{fig:dsig_dxf_kmr_scales}
we show the dependence of the differential cross section on the value
of the scale. There is a sizeable effect (about a factor of two) on
the cross section (see also Table 1 with integrated cross sections).
For $x_F \sim$ 0, the smaller $\mu^2$ the bigger the cross section.


\begin{figure}[!h]    %
\includegraphics[width=0.55\textwidth]{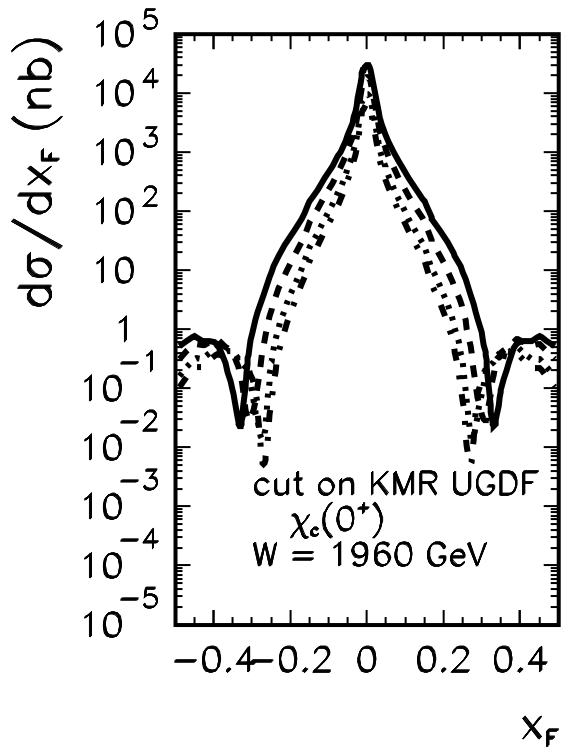}
   \caption{\label{fig:dsig_dxf_kmr_cuts}
   \small  The effect of the cut on low gluon transverse momenta
on distribution in Feynman $x_F$ in the KMR approach
(KMR UGDF, KMR vertex). The solid line is for $Q_{cut}^2$ = 0.26
GeV$^2$, the dashed line for $Q_{cut}^2$ = 0.5 GeV$^2$,
the dotted line for $Q_{cut}^2$ = 0.8 GeV$^2$ and dash-dotted line
for $Q_{cut}^2$ = 1.0 GeV$^2$.
The calculation was done for the Tevatron energy W = 1960 GeV.}
\end{figure}

In order to demonstrate sensitivity to the nonperturbative region
of small transverse momenta in Fig.\ref{fig:dsig_dxf_kmr_cuts} we show
results with the KMR UGDF cut off for small values of the effective
gluon transverse momenta $Q_t^2 < Q_{cut}^2$. One observes
a rather strong dependence on the value of the cut-off parameter.
The bigger the value of the cut-off parameter the smaller the cross
section. The results for the integrated cross section are
summarized in Table I.
Even for large values of the cut-off parameter ($Q_{cut}^2 \sim$ 1
GeV$^2$) sizeable cross sections are obtained.


\begin{figure}[!h]    %
\includegraphics[width=0.55\textwidth]{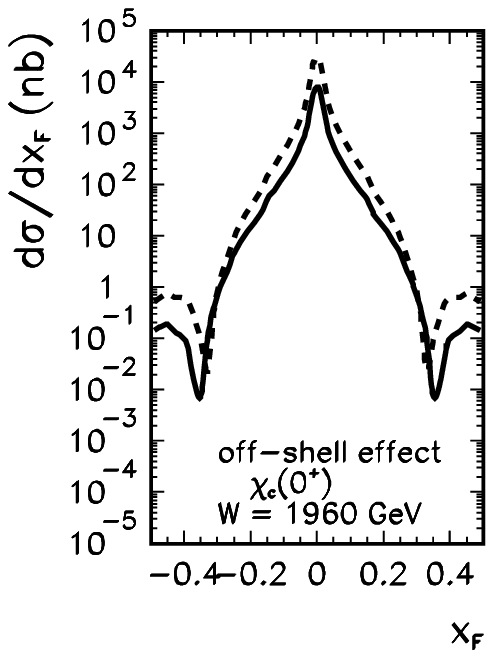}
   \caption{\label{fig:dsig_dxf_kmr_off_shell}
   \small  Off-shell effect on distribution in Feynman $x_F$ in the
 KMR approach. The dashed line is for on-shell case and solid line is
for off-shell case.
The calculation was done for the Tevatron energy W = 1960 GeV.}
\end{figure}

The estimate in Ref.\cite{KMR-chi} was done assuming that the gluons
in the $g g \to \chi_c(0^+)$ vertex are on mass shell, which is
exact only in the infinite meson mass limit. In this paper
(see subsection IIB) we take into account the effect of gluon
virtualities. In Fig.\ref{fig:dsig_dxf_kmr_off_shell} we show the role
of the off-shell effects for the KMR UGDF. The off-shell effect leads
to a reduction of the cross section by a factor of 2 -- 5.


Finally we show the influence of the off-shell effects on
azimuthal-angle correlation function. The off-shell effect in
the matrix element reduces the cross section but does not change
its shape. The shape of the distribution requires an extra comment.
The distribution in azimuthal angle is very different
from $(1 + \cos(2 \Phi))$ as expected for one-step fusion
(e.g. photon-photon, pomeron-pomeron fusion) to be discussed in detail
in the next subsection.


\begin{figure}[!h]    %
\includegraphics[width=0.55\textwidth]{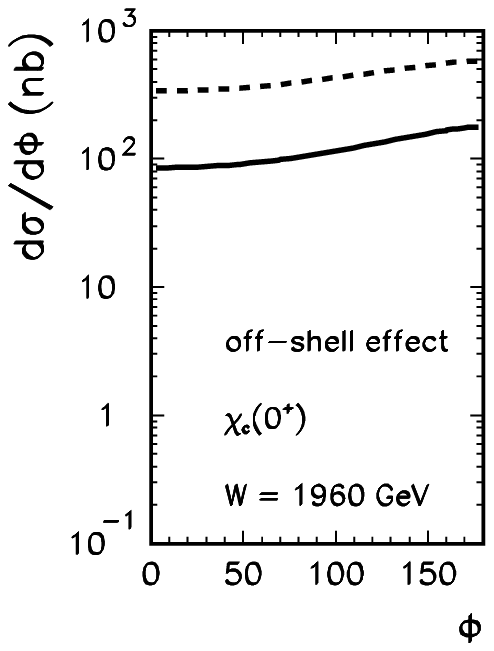}
   \caption{\label{fig:dsig_dphi_kmr}
   \small  Off-shell effect on distribution on azimuthal correlation
function in the KMR approach. The dashed line is for on-shell case and
solid line is for off-shell case.
The calculation was done for the Tevatron energy W = 1960 GeV.}
\end{figure}


\subsection{Our approach}

\begin{figure}[!h]    %
\includegraphics[width=0.4\textwidth]{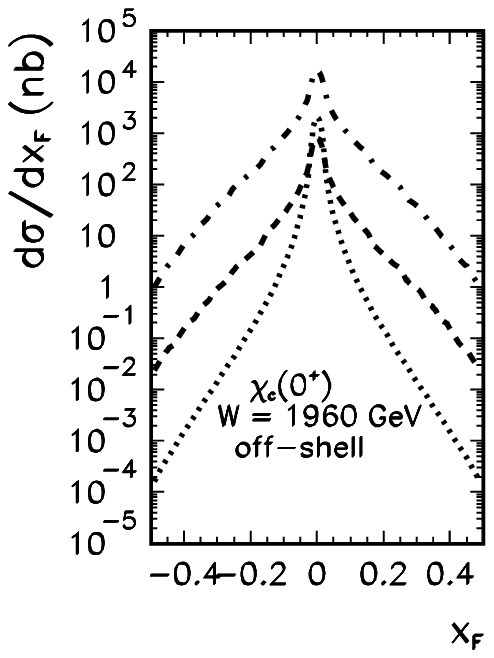}
\includegraphics[width=0.4\textwidth]{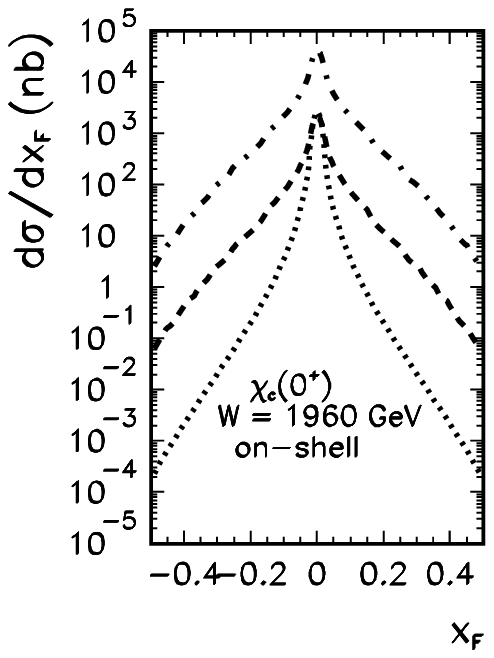}
   \caption{\label{fig:dsig_dxf}
   \small  Distribution in Feynman $x_F$ for different UGDFs
(KL--dashed, GBW--dotted, BFKL--dash-dotted)
for off-shell (left panel) and on-shell (right panel) matrix element.
The calculation was done for the Tevatron energy W = 1960 GeV.}
\end{figure}

In Fig.\ref{fig:dsig_dxf} we present our distributions
of the $\chi_c(0^+)$ mesons in the Feynman variable $x_F$. We present
results obtained with different UGDFs.
Characteristic for central diffractive production all distributions
peak at $x_F \approx$ 0.
Although all UGDFs give a similar quality description of
the low-$x$ HERA data for the $F_2$ structure function,
they give quite different longitudinal momentum distributions of
$\chi_c(0^+)$ at the Tevatron energy $W$ = 1960 GeV.
The UGDFs which take into account saturation effects (GBW, KL) give much
lower cross section than the BFKL UGDF (dash-dotted line).
Similar as in Ref.\cite{SPT07} rather small values of $x$'s in formula
(\ref{ampl}) come into the game here. Therefore the process considered here
would help, at least in principle, to constrain the poorly known UGDFs.


\begin{figure}[!h]    %
\includegraphics[width=0.55\textwidth]{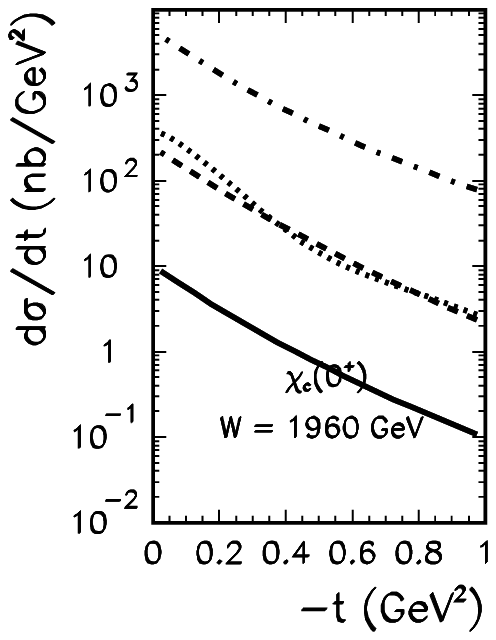}
   \caption{\label{fig:dsig_dt}
   \small  Distribution in $t = t_1 = t_2$ for different UGDFs.
The notation here is the same as in Fig.\ref{fig:dsig_dxf}.
The solid line is for the Gaussian distributions with
$\sigma_0$ = 1.0 GeV and the second choice of the scales in
Eq.(\ref{Gaussian_scales}).
}
\end{figure}

In Fig.\ref{fig:dsig_dt} we show distributions in the square of the
four-momentum transfers ($t_1$ or $t_2$) in the nucleons lines.
Because they are identical we shall denote them $d \sigma / dt$ for
brevity.
The distributions shown in the figure are peaked at small values of
$t_1$ or $t_2$. Slightly different slopes are obtained with different
UGDFs. This demonstrates how purely known are UGDFs at present.
The measurement of such distributions requires measuring forward
protons (or antiprotons).
In the case no measurement of forward nucleons is possible one should
study events with low multiplicity (only particles from the decay of
$\chi_c(0^+)$). This can be either pairs of pions or pairs of kaons
or pairs of photons or $J/\psi + \gamma$. Unfortunately branching ratios to
these channels are rather low \cite{PDG06}.


\begin{figure}[!h]    %
\includegraphics[width=0.55\textwidth]{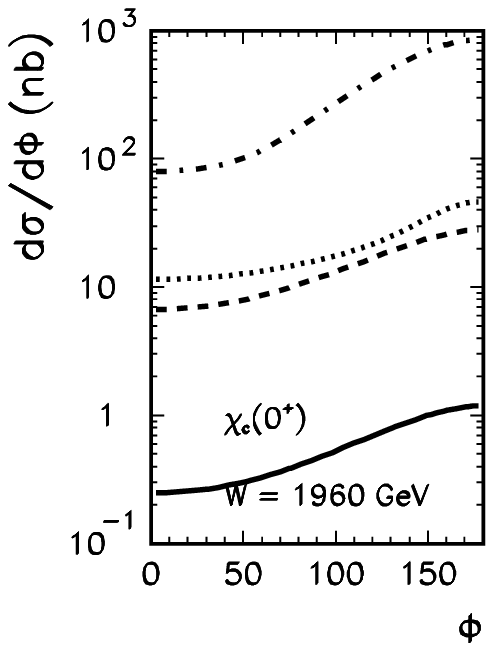}
   \caption{\label{fig:dsig_dphi}
   \small  Distribution in Feynman relative azimuthal angle
for different UGDFs
The notation here is the same as in Fig.\ref{fig:dsig_dxf}.
The solid line is for the Gaussian distributions with
$\sigma_0$ = 1.0 GeV and the second choice of the scales in
Eq.(\ref{Gaussian_scales}).
}
\end{figure}

In Fig.\ref{fig:dsig_dphi} we show azimuthal angular correlations
of the outgoing nucleons. Measuring such distribution experimentally
requires identification of both nucleons in very forward/backward
directions. This is not possible with the present Tevatron apparatus.
We hope such a measurement will be possible with the final LHC
instrumentation. All distributions shown in Fig.\ref{fig:dsig_dphi}
are peaked at $\Phi \sim$ 180$^0$ i.e. for the back-to-back kinematics.
The fact that the distributions are not simple
functions (sin$\Phi$, cos$\Phi$) of the relative azimuthal angle
between outgoing nucleons is due to the loop integral
in Eq.(\ref{ampl}) which destroys the dependence one would obtained
with single fusion of well defined objects (mesons or reggeons).


\begin{figure}[!h]    %
\includegraphics[width=0.4\textwidth]{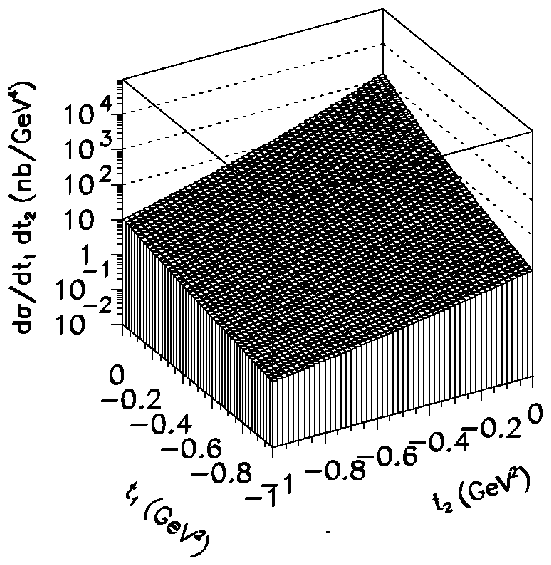}
\includegraphics[width=0.4\textwidth]{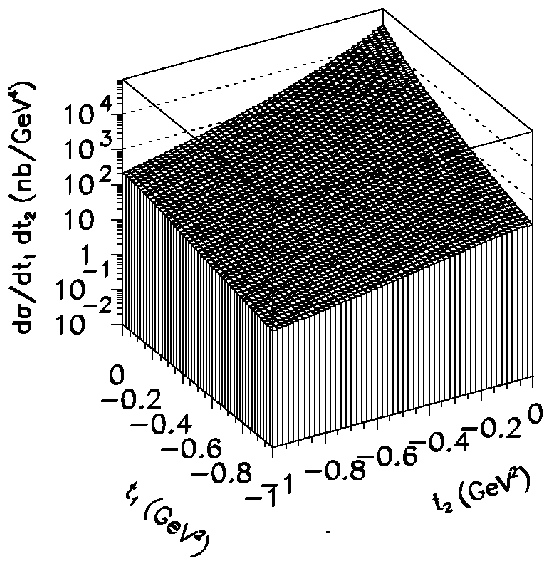}
\includegraphics[width=0.4\textwidth]{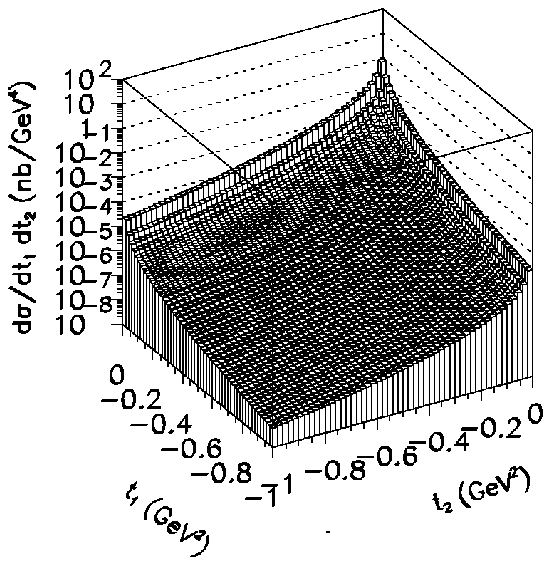}
\includegraphics[width=0.4\textwidth]{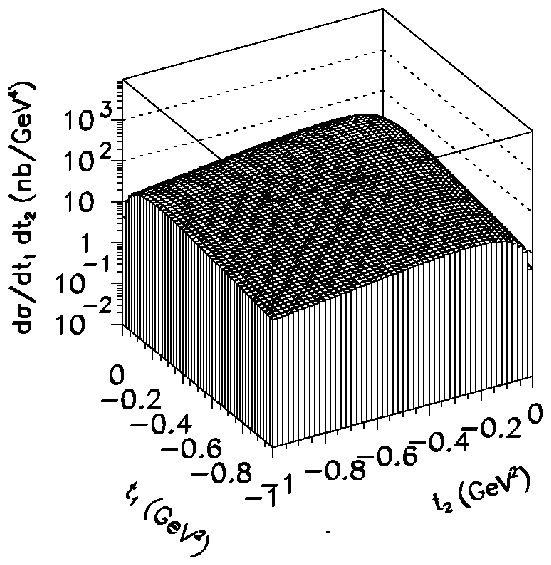}
   \caption{\label{fig:maps_t1t2}
   \small  Two-dimensional maps in $t_1$ and $t_2$ for
KL UGDF (top left) and BFKL UGDF (top right) as well as
for two-photon fusion (bottom left) and two-pomeron fusion
with dipole form factors (bottom right).}
\end{figure}

Another type of proton-antiproton correlation is shown
in Fig.\ref{fig:maps_t1t2}. We show results for KL and BFKL UGDFs
as well as for photon-photon fusion. Finally we show the two-dimensional
distribution in $t_1$ and $t_2$ for the ``fusion'' of two vector
pomerons. The ($t_1,t_2$) distribution obtained in the photon-photon fusion
mechanism differs qualitatively from the all other distributions.
One can see a strong enhancement of the cross section when
$t_1 \to$ 0 or $t_2 \to$ 0.
Also the shape of the two-dimensional spectrum obtained for the fusion
of phenomenological vector pomerons differs from those obtained in
the QCD-inspired KMR mechanism (please note minimum of the cross section
when $t_1 \to$ 0 and/or $t_2 \to$ 0).


\begin{figure}[!h]    %
\includegraphics[width=0.4\textwidth]{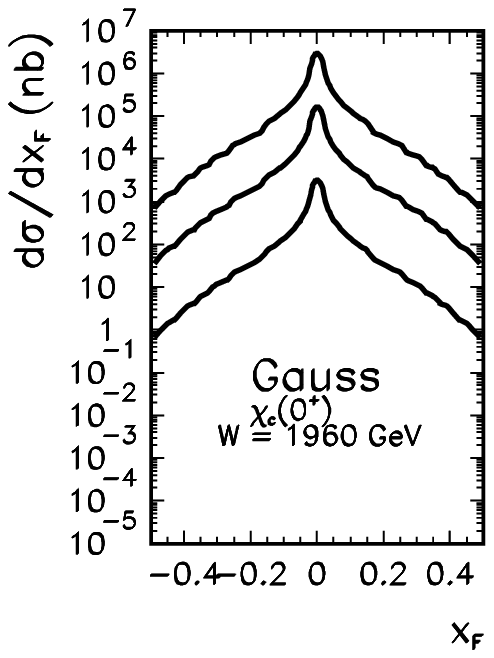}
\includegraphics[width=0.4\textwidth]{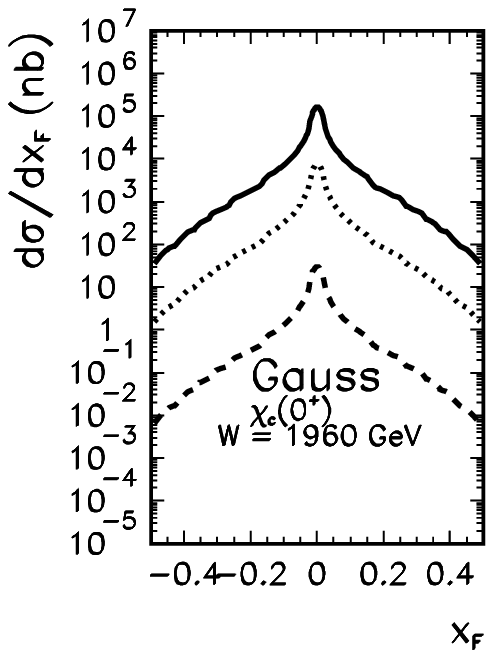}
   \caption{\label{fig:dsig_dxf_gauss}
   \small  Distribution in $x_F$ of the $\chi_c(0^+)$ meson for the Gaussian
UGDF. In the left panel different values of the parameter
$\sigma_0$ = 0.5, 1, 2 GeV and $\mu_0^2 = \mu^2 = M_{\chi}^2$.
In the right panel $\sigma_0$ = 1 GeV and we have chosen different
scales in Eq.(\ref{Gaussian_scales}): 1: solid, 2: dashed and 3: dotted.
}
\end{figure}

In order to demonstrate the role of the gluon transverse momenta
in UGDFs
in Fig.\ref{fig:dsig_dxf_gauss} we show distributions in $x_F$ for
the Gaussian UGDF for different values of the smearing parameter $\sigma_0$.
We wish to notice here that all such UGDFs correspond to identical
integrated GDFs. The smaller value of $\sigma_0$ the larger the cross
section. This demonstrates that the main contributions to the cross
section come from the region of very small transverse momenta of the
t-channel gluons (see Fig.\ref{fig:kinematics_qcd}).
This is cleary the region where nonperturbative effects
are dominant.


\begin{figure}[!h]    %
\includegraphics[width=0.55\textwidth]{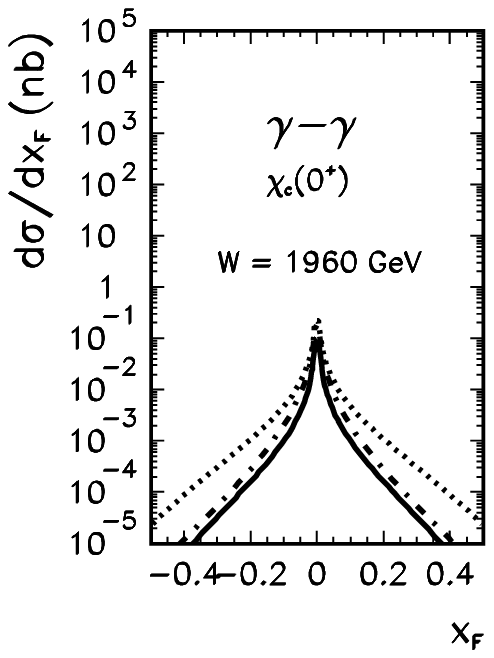}
   \caption{\label{fig:dsig_dxf_gamgam}
   \small  Distribution in $x_F$ for the two-photon fusion.
The solid line represents result obtained
with Eq.(\ref{gamma-gamma-chic-amp}).
The dotted and dashed dotted lines correspond to the EPA approach
with different flux factors as described in the text.}
\end{figure}

Let us turn now to the second mechanism of the exclusive $\chi_c(0^+)$
production -- the two-photon fusion mechanism sketched in Fig.\ref{fig:gamgam}.
In Fig.\ref{fig:dsig_dxf_gamgam} we show
corresponding distribution in the Feynman variable $x_F$.
For comparison we show also
the EPA result (dotted line) discussed in subsection IIIB.

\begin{figure}[!h]    %
\includegraphics[width=0.55\textwidth]{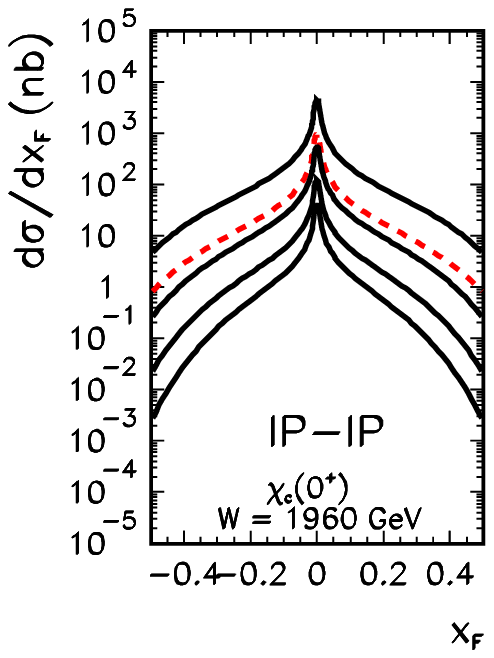}
   \caption{\label{fig:dsig_dxf_pompom}
   \small
Distribution in $x_F$ for the two-pomeron fusion for different values
of the slope parameter B = 5, 10, 15, 20 GeV$^{-2}$ (from top to bottom)
of the exponential form factor. The dashed line is obtained with the dipole
(electromagnetic) form factor.}
\end{figure}

For pedagogical purpose in
Fig.\ref{fig:dsig_dxf_pompom} we show also the result of
the calculation with vector-like pomerons as described in section IV.
Rather smaller cross section is obtained compared to the more QCD-inspired
calculation with UGDFs.
The cross section depends strongly on the value of the
parameters of the form factor describing coupling of the phenomenological
pomeron to nucleons. The value of $B$ = 15-20 GeV$^{-2}$ is preferred
from the proton-proton or proton-antiproton elastic scattering phenomenology.
Compared to the calculation with UGDFs presented in
Fig.\ref{fig:dsig_dxf} the distribution around $x_F$ = 0 is much broader
here.
The thinner distribution in the case of the $k_t$-factorization approach
is due to the $x$-dependence of the UGDFs entering the basic formula.
The values of x's ($x_1, x_2$, etc.) in the off-diagonal UGDFs change
quickly with the Feynman variable $x_F$ which makes the peak at $x_F$ =
0 much thinner than for the phenomenological pomeron exchange discussed here.


\begin{figure}[!h]    %
\includegraphics[width=0.55\textwidth]{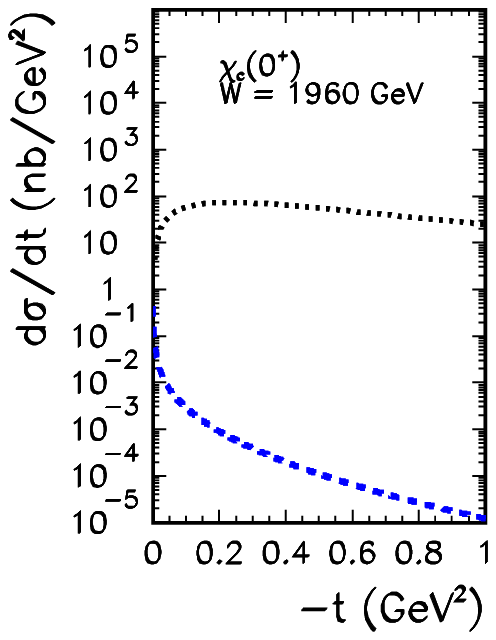}
   \caption{\label{fig:dsig_dt_gamgam_vs_pompom}
   \small  Distribution in $t = t_1 = t_2$ for the two-photon (dashed,
   blue on-line) and two-pomeron (dotted) fusion.
   Compare these distributions with those obtained in the QCD approach
   in Fig.\ref{fig:dsig_dt}.
}
\end{figure}

Let us return for a while to $t$-distributions.
In Fig.\ref{fig:dsig_dt_gamgam_vs_pompom} we show single distributions
for photon-photon and Pomeron-Pomeron fusion mechanisms
in $t = t_1 = t_2$ (relevant for one-nucleon tagged case) obtained by
projecting the two-dimensional distributions shown in
Fig.\ref{fig:maps_t1t2}.
The electromagnetic component peaks at very small values of $t$ due
to the photon propagators.

The results for the diffractive mechanism depend strongly
on the details of the calculation. Here we show only one distribution
of diffractive component for easy reference. Although the diffractive
component is subjected to much stronger absorption effects than
the electromagnetic one, it is clear that the diffractive component dominates.


\begin{figure}[!h]    %
\includegraphics[width=0.55\textwidth]{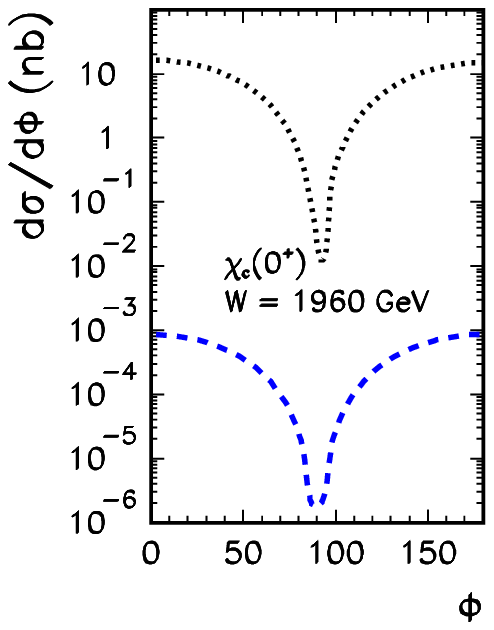}
   \caption{\label{fig:dsig_dphi_gamgam_vs_pompom}
   \small  Distribution in relative azimuthal angle $\Phi$ for
   the two-photon (dashed,
   blue on-line) and two-pomeron (dotted) fusion.
   Compare these distributions with those obtained in the QCD approach
   in Fig.\ref{fig:dsig_dphi}.
}
\end{figure}

In Fig.\ref{fig:dsig_dphi_gamgam_vs_pompom} we show differential
distribution in relative azimuthal angle for $\gamma^* \gamma^*$ and
Pomeron-Pomeron fusion mechanisms. One sees a typical
$(1 + \cos(2 \Phi))$ dependence characteristic for one-step exchanges.
These distributions are very different from those shown in
Fig.\ref{fig:dsig_dphi} for pQCD diffraction where the underlying
mechanism is more complicated due to a two-step nature of the process
(see Fig.\ref{fig:kinematics_qcd}).


Above we have calculated only bare distributions.
Those are subjected to absorption effects.
The absorption effects are usually included by multiplying the bare
distributions by a soft gap survival probability.
The gap survival probability was estimated to $S^2(Tevatron) \approx$ 0.05
and $S^2(LHC) \approx$ 0.025 \cite{KMR-chi}. Absorption leads therefore
to a large reduction of the bare cross section.
In principle, the absorption effects may modify also shapes
of differential distributions. This requires a further detailed analysis
which goes beyond the scope of the present paper.

As discussed in this paper there are huge uncertainties in estimating
the cross section for exclusive $\chi_c(0^+)$ production, much larger e.g.
than for the exlusive $J/\psi$ production where the corresponding
amplitude can be related to the amplitude of $J/\psi$ photoproduction
in $e p$ collisions \cite{SS07}.
Only a real experiment may shed more light on the dynamics of
$\chi_c(0^+)$ production mechanism. Measuring centrally produced
$\chi_c(0^+)$, one forward nucleon (proton or antiproton)
and imposing a condition of rapidity gap in the second hemisphere
should allow, at least in principle, such a measurement.

In Table I we have combined the cross section integrated over the whole
available phase space for exclusive production of the $\chi_c(0^+)$
mesons.

\begin{table}

\caption{
Integrated cross section (in nb) for exclusive $\chi_c(0^+)$ production
for the Tevatron energy W = 1960 GeV in various possible theoretical
prescriptions. The number for the Gaussian UGDF
refers to the number in Eq.(\ref{Gaussian_scales}).
}

\begin{center}

\begin{tabular}{|c|c|c|c|}
\hline
UGDF, details &$\qquad\sigma_{tot}\qquad$& $\quad
S^2\,\sigma_{tot}\quad$& $\;\;\mathrm{BR}\,S^2\,\sigma_{tot}\;\;$\\
\hline
1. KMR, $Q_{cut}^2=0.26$ GeV$^2$, on-shell &          &          &          \\
 vertex, min prescription for $Q_t^2$      &0.1357(+4)&0.1357(+3)&0.1357(+1)\\
2. KMR, $Q_{cut}^2=0.26$ GeV$^2$, on-shell &          &          &          \\
 vertex, max prescription for $Q_t^2$      &0.9628(+4)&0.9628(+3)&0.9628(+1)\\
3. KMR, $Q_{cut}^2=0.26$ GeV$^2$, off-shell&          &          &          \\
 vertex, min prescription for $Q_t^2$      &0.3720(+4)&0.3720(+3)&0.3720(+1)\\
4. KMR, $Q_{cut}^2=0.5$ GeV$^2$, on-shell  &          &          &          \\
 vertex, min prescription for $Q_t^2$      &0.8227(+3)&0.8227(+2)&0.8227(+0)\\
5. KMR, $Q_{cut}^2=0.8$ GeV$^2$, on-shell  &          &          &          \\
 vertex, min prescription for $Q_t^2$      &0.4124(+3)&0.4124(+2)&0.4124(+0)\\
6. KMR, $Q_{cut}^2=1.0$ GeV$^2$, on-shell  &          &          &          \\
 vertex, min prescription for $Q_t^2$      &0.2745(+3)&0.2745(+2)&0.2745(+0)\\
\hline
4. KL, on-shell vertex              &0.1180(+3)&0.1180(+2)&0.1180(+0)\\
5. KL, off-shell vertex             &0.5231(+2)&0.5231(+1)&0.5231(-1)\\
\hline
6. GBW, on-shell vertex             &0.8514(+2)&0.8514(+1)&0.8514(-1)\\
7. GBW, off-shell vertex            &0.1590(+3)&0.1590(+2)&0.1590(+0)\\
\hline
8. BFKL, on-shell vertex            &0.2603(+4)&0.2603(+3)&0.2603(+1)\\
9. BFKL, off-shell vertex           &0.1125(+4)&0.1125(+3)&0.1125(+1)\\
\hline
10. Gauss, $\sigma_0=0.5$ GeV,      &          &          &          \\
off-shell vertex, scales (2)        &0.2141(+2)&0.2141(+1)&0.2141(-1)\\
11. Gauss, $\sigma_0=1.0$ GeV,      &          &          &          \\
off-shell vertex, scales (2)        &0.1811(+1)&0.1811(+0)&0.1811(-2)\\
\hline
\end{tabular}

\end{center}

\end{table}


In Table 1 we have assumed $W$ = 1960 GeV, $S^2$ = 0.1 and
BR$(\chi_c(0^+)\to J/\psi+\gamma) = 0.01$.
The recently measured value of the branching ratio
is 1.3 $\pm$ 0.11 \% \cite{PDG06}.
The last column shows possible contribution of the
$\chi_c(0^+) \to J/\psi + \gamma$ decay to the exclusive production
of $J/\psi$ if the soft photon cannot be correctly identified.
We get typically 0.1-5 nb, i.e. much less than the direct $J/\psi$
photoproduction \cite{SS07}. It would be interesting to calculate
the contribution of $\chi_c(1^+) \to J/\psi \gamma$ and
$\chi_c(2^+) \to J/\Psi \gamma$, where the corresponding branching
ratios are order of magnitude larger. On the other hand, within the
approximations made in Ref.\cite{Yuan01} the cross sections
for diffractive production of $\chi_c(1^+)$ and $\chi_c(2^+)$ vanish.
It would be interesting to go beyond the approximations of
Ref.\cite{Yuan01}.

The exclusive production of $\chi_c$ has been reported recently
by the CDF collaboration \cite{CDF_limit} with the upper limit for
the cross section
\begin{equation}
\sigma_{exc} ( p \bar p \to p + J/\psi + \gamma + \bar p)
< 49 \pm 18 (stat) \pm 39 (sys) \; pb \; ,
\label{CDF_limit}
\end{equation}
within the CDF experimental cuts.
These numbers cannot be directly compared with our results in Table 1
that do not include the experimental cuts. Such a comparison requires
a Monte Carlo type analysis as was done in Ref.\cite{RRABP07}.


\subsection{Energy dependence}

\begin{figure}[!h]    %
\includegraphics[width=0.4\textwidth]{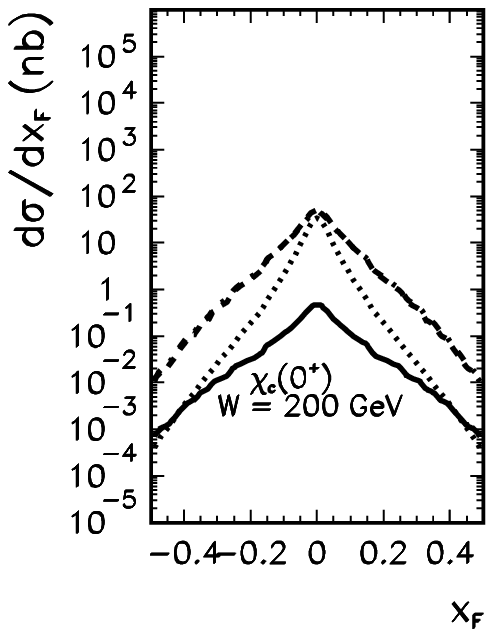}
\includegraphics[width=0.4\textwidth]{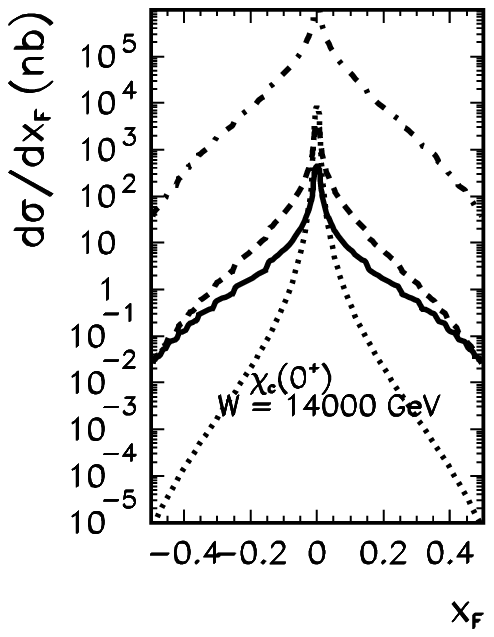}
   \caption{\label{fig:dsig_dxf_energy_dependence}
   \small Distribution in $x_F$ of $\chi_c(0^+)$ meson for different
UGDFs. In the left panel we show results for RHIC and in the right panel
for LHC. The meaning of the lines is the same as for the Tevatron energy.
}
\end{figure}

Up to now we have concentrated on Tevatron energy. We expect some
results for exclusive $\chi_c$ production in not too distant future.
It would be also interesting to measure exclusive $\chi_c$ production
at different energies. The obvious choices are RHIC and LHC in the near
future.
In Fig.\ref{fig:dsig_dxf_energy_dependence} we show distributions
in $x_F$ for different UGDFs for RHIC, W = 200 GeV (left panel) and
LHC, W = 14000 GeV (right panel). Compared to Tevatron the distributions
in $x_F$ for RHIC are wider and distribution for LHC are more narrow,
concentrated around $x_F$ = 0. While at RHIC energy different UGDFs
give relatively similar results, at LHC energy there is a
difference of several orders of magnitude between results obtained with
different UGDFs. Therefore a measurement at LHC should clearly
select the best distribution.

In Table II we have collected total cross sections for selected UGDFs
at RHIC, Tevatron and LHC. Comparing results for the three different
energies we see that different UGDFs predict completely different
energy dependence. While BFKL predicts a strong growth of the cross
section with the collision energy, the saturation models (KL, GBW)
predict much slower rise with the collison energy.
Therefore measuring exclusive $\chi_c(0^+)$ production at three different
energies would be useful to pin down underlying dynamics.


\begin{table}

\caption{
Integrated cross section $\sigma_{tot}$ (in nb) for exclusive
$\chi_c(0^+)$ production at different energies. The number for the Gaussian UGDF
refers to the number in Eq.(\ref{Gaussian_scales}).}

\begin{center}
\begin{tabular}{|c|c|c|c|}
\hline
UGDF &$\qquad$RHIC$\qquad$&$\qquad$Tevatron$\qquad$& $\qquad$LHC$\qquad$\\
\hline
Kl                  &0.6430(+1)&0.5231(+2)&0.1090(+3)\\
GBW                 &0.2830(+1)&0.1590(+3)&0.1413(+3)\\
BFKl                &0.6140(+1)&0.1125(+4)&0.6306(+5)\\
\hline
Gauss,
$\sigma_0=1.0$ GeV, &          &          &          \\
scales (2)          &0.7126(-1)&0.1811(+1)&0.1428(+2)\\
\hline
\end{tabular}
\end{center}
\end{table}



\section{Discussion and Conclusions}

In the present paper we have discussed in detail the exclusive
production of the $\chi_c(0^+)$ meson in proton-antiproton and
proton-proton collisions.
We have considered both diffractive and purely electromagnetic
mechanisms. Many differential distributions
have been discussed for the first time in the literature.

The diffractive component was calculated in the Khoze-Martin-Ryskin
approach. Compared to the original KMR calculation we have taken into
account the off-shellness of the gluons in the corresponding diagram.
The corresponding matrix element was calculated. We find that
the inclusion of the gluon virtualities reduces the cross section
by a factor of 2 -- 5, depending on kinematical region and on UGDFs.

We have discussed uncertainties in the KMR approach related to
the treatment of the nonperturbative region as well as related to
the choice of the scale in their skewed unintegrated distributions.
This gives an uncertainty of a factor 2 -- 3. 

We get similar integrated cross section as in Ref.\cite{KMR-chi} if we
take the on-shell approximation for the vertex, make the same choice of
scales, use the same integrated gluon distributions, etc.
Summarizing, we find rather large uncertainties in the approach.

Many other UGDFs from the literature were also used to calculate
distributions in the Feynman variable $x_F$, in squares of
the four-momentum transfers
($t_1$ and $t_2$) in the nucleon lines and in relative azimuthal angle
between outgoing protons. Also correlations in $t_1$ and $t_2$ have been
analyzed.
The results depend strongly on the choice of UGDF.
This is related to a particular sensitivity of the cross section
for the reaction under consideration to the nonperturbative region of
very small gluon transverse momenta. Therfore a measurement of
the differential distributions would be very helpful to test
the unintegrated distributions in this region.
At RHIC one could measure the $\pi^+ \pi^-$ and/or $K^+K^-$ decay
channels. Probably ALICE could use a similar method.
At Tevatron rather $\gamma \gamma$ or $\gamma J/\psi$
channels seem preferable.

In the present paper we have calculated only bare distributions.
Those are subjected to absorption effects.
The absorption effects are usually included by multiplying the bare
distributions by a soft gap survival probability.
The survival probability was estimate as $S^2(Tevatron) \approx$ 0.05
and $S^2(LHC) \approx$ 0.025 \cite{KMR-chi}. Absorption leads therefore
to a large reduction of the bare cross section.
In principle the absorption effects may modify also shapes
of differential distributions \cite{SS07}. This requires a further
detailed analysis which goes, however, beyond the scope of the present paper.

For completeness we have calculated also the cross section in a more
phenomenological approach with Regge-type pomeron-pomeron fusion.
Cross sections of the same order of magnitude as for the QCD approach
are obtained. However, the azimuthal angle correlation functions
for both approaches are very different.

We have also calculated differential distributions for the photon-photon
fusion. This contribution turned out to be rather small (a fraction
of nb). The differential distributions have quite different shapes
compared to the diffractive component. Compared to the diffractive
component it is peaked at extremely small values of $t_1$ and/or $t_2$.
Also the azimuthal angle correlations pattern is very different:
$\cos \Phi$ for the $\gamma \gamma$ fusion
and a more complicated shape for the diffractive component
due to the loop integration in the formula for the amplitude.

Recently, there is an interest at Tevatron to study exclusive production
of $J/\psi$ meson.
This is because of a possibility to study $J/\psi$ photoproduction
at large energies \cite{SS07} and due to a potential search
for odderon exchange \cite{BMSC07}. Because of its decay channel
$\chi_c(0^+) \to \gamma J/\psi$ and difficulties in identifying soft
photons the $\chi_c(0^+)$ decays may contribute to the ``exclusive''
production of $J/\Psi$ making the measurement of $J/\Psi$ photoproduction
and/or the discovery of the odderon exchange difficult.
A comparison of theoretical cross sections for the three mentioned
reactions would be therefore very useful in this context.
Our estimates for Tevatron energies, including soft survival
probability, give (20 -- 200) nb times 
$BR(\chi_c(0^+) \to \gamma J/\psi)$ = 0.013 which
gives the corresponding cross section rather less than 1 nb compared
to about 15 nb for the $J/\psi$ photoproduction \cite{SS07}.

We have made also calculation of $d \sigma / d x_F$ distributions
and total cross sections for RHIC and LHC. Comparing these results
and those for the Tevatron leads to the conclusion that only measurements
for different energies may help to disentangle underlying QCD
dynamics. In principle, such measurements could even explore the onset 
of QCD saturation which is not so easy to be discovered in inclusive
reactions.

\vskip 0.5cm

{\bf Acknowledgements} This work was partially supported by the
grant of the Polish Ministry of Scientific Research and Information
Technology number 1 P03B 028 28, the Russian Foundation for
Fundamental Research, grants No. 06-02-16215 and No. 07-02-91557,
and the Bogoliubov-Infeld Programme 2007.



\end{document}